\documentclass[3p,english,preprint]{elsarticle}

\usepackage{ecrc}

\volume{00}

\firstpage{1}

\journalname{Electrochimica Acta}

\runauth{Elshad Allahyarov et al}

\jid{elca}

\jnltitlelogo{Electrochimica Acta}

\usepackage{amssymb}

\usepackage[figuresright]{rotating}

\usepackage{epsfig}
\usepackage{color}
\usepackage{float}
\usepackage{graphicx}
\usepackage{epstopdf}
\usepackage{babel}
\usepackage{epsfig}
\newcommand{\need}[1]{\textcolor{black}{#1}}
\newcommand{\mod}[1]{\textcolor{black}{#1}}

\makeatother

\begin{document}

\begin{frontmatter}



\dochead{Invited article for Electrochimica Acta,  guest edited by Michael Eikerling}

\title{Simulation Study of Ion Diffusion in Charged Nanopores with Anchored Terminal Groups}


\author[1,2,3]{Elshad Allahyarov}
 \ead{elshad.allakhyarov@case.edu}

\author[1]{ Hartmut L{\"o}wen}

\author[4]{  Philip L. Taylor}

\address[1]{Institut f\"ur Theoretische Physik II: Weiche Materie, 
Heinrich-Heine Universit\"at  D\"usseldorf,  
  Universit\"atstrasse 1, 40225 D\"usseldorf, Germany}     
\address[2]{ Department of Macromolecular Science and Engineering, Case Western Reserve   
University, Cleveland, Ohio 44106-7202, United States}
\address[3]{Theoretical Department, Joint Institute for High Temperatures, Russian Academy of  
Sciences (IVTAN), 13/19 Izhorskaya street, Moscow 125412, Russia}
\address[4]{ Department of Physics, Case Western Reserve University,
  Cleveland, Ohio 44106-7079, United States}

\begin{abstract}

We present  coarse-grained simulation results for  enhanced ion diffusion in a charged nanopore grafted 
with ionomer sidechains. The pore surface is hydrophobic and its diameter is varied from 2.0 nm to 3.7 nm. 
The sidechains have from  2 to 16 monomers (united atom units) and contain sulfonate terminal groups. 
 Our simulation results indicate a strong dependence of the ion diffusion along the pore axis on the pore parameters.
In the case of short sidechains and large pores  the ions mostly occupy the pore wall area, where their  
distribution is strongly disturbed by their host  sulfonates.
In the case of short sidechains and narrow pores, 
the mobility of ions is strongly affected  by the structuring and polarization effects  of the  water molecules. 
In the case of long sidechains, and  when the sidechain sulfonates reach the pore center,   
a radial charge separation occurs in the pore. Such charge separation  suppresses the ion diffusion along the pore axis. 
 An enhanced ion diffusion was found in  the pores grafted with 
 medium-size sidechains provided that 
the  ions do not enter the central pore area, and the water is less structured around the ions and sulfonates. 
In this case,  the 3D density of the ions has 
a hollow-cylinder type shape with a smooth and uninterrupted 
surface. We found that the  maximal ion diffusion  has a linear dependence on the 
number of sidechain monomers.  It is suggested that 
the maximal ion diffusion along the pore axis is attained if the 
effective length of the sidechain extension into the pore center
 (measured as twice the gyration radius of the sidechain 
with the Flory exponent 1/4) is about 1/3 of the pore radius.

\end{abstract}


\end{frontmatter}


\section{Introduction}
\label{intro}

Over the past decade there has been  a growing interest in  ion 
transport phenomena in restricted geometries 
and porous networks. The development of many applications   relies
 on a fundamental understanding of the diffusion properties
 of particles in cylindrical pores.  
For example, transport properties of the catalyst layer of polymer electrolyte fuel cells 
 are largely regulated by the electrostatic interaction of the ion with the charges  on the
cylindrical pore in the Pt electrode 
\cite{eikerling-2016-pt-water-ionomer,eikerling-2014-pt-nanochannels,eikerling-2011-pre-in-ionomer-catalyst-layer}.
The charge storage in conducting narrow nanopores  relies on the voltage-controlled 
 accumulation of ions in a narrow metallic nanopore
\cite{kornyshev-2014-charge-storage-single-file-pore,kornyshev-2014-charge-storage,zhang-2012-nanoporous-metals}. 
Charged nanopores are used for the  partitioning of ions and proteins 
\cite{rohani-2010-protein-ultrafiltration-in-charged-pore,armstrong-2013-ion selection-with-different-charges-and-mobilities-pores-grafted-charged-sidechains,stroeve-2014-protein-transport-nanopore}, 
for electrolyte nanofiltration \cite{ryzhkov-2016-nanopore}, 
 and for ion-current rectification 
\cite{oeffelen-2015-ion-current-nanopore}. 
 A  charged-pore model also has been adopted as a 
convenient study model for   
 water sorption \cite{eikerling-2011-pore-swelling-in-ionomer} and  
structural and kinetic characteristics of ion conductivity
in fuel cell ionomers   
\cite{eikerling-kornyshev-2001,paddison-2003-review,paul-paddison-2005-pore}, 
for the self-diffusion of ions in nanoporous media
\cite{jardat-2012-ion-diffusion,malgaretti-2016-charged-ion-nanopore}, 
for the permeability of the ion channels to water and ions \cite{dzubella-2005}, 
and 
for the electro-osmotic flow of ionic solutions in charged channels \cite{kim-2006}.
Nanopores with anchored ionizable surface groups have been used for  
 advanced membrane separations
\cite{szymzyk-2010charged-nanopore-ion-rejection}, 
for ion selectivity  
\cite{gassara-2015-nanofiltration-ion-grafted-charged-polymer},
  and for the enhanced transport of proteins
\cite{basconi-2015-polymer-graft-protein-transport,basconi-2014-polymer-graft-protein-transport}.
They also serve as a model for 
the ion diffusion in  cylindrical pores of the  membrane-ionomer interface
\cite{paddison-2013}, and
for the permeability of the ionomer pores to specific ions 
\cite{yang-pintauro-2004,pintauro-1995}.

Ongoing research on new polyelectrolyte membranes (PEM) for 
fuel cell applications also focuses on the ion diffusion properties of porous networks
\cite{mauritz2004,gebel-2000,ioselevich,Gierke2}.
Ionomer chains of the PEM  are mostly composed of hydrophobic inert
 backbones grafted with pendant side  chains that are terminated by anionic headgroups.
When hydrated, the terminal group ions dissociate and the sulfonates self 
assemble into connected clusters, creating hydrophilic pathways inside 
the hydrophobic backbone matrix. 
For effective ion transport through the membrane, the PEM needs to 
be  hydrated and swollen in order to form a connected network  of hydrophilic pathways.  
The hydration water, however,  makes the 
PEM vulnerable to icing, boiling and water evaporation  at low and high temperatures
and  
to loss of its elastic properties. In this sense, designing  new
ionomers with low water uptake and, at the same time, with acceptable ionic conductivity, 
is the goal of many ongoing activities in developing advanced  fuel cell membranes.  

One of the possible ways to proceed in this direction is the  concept of 
a matrix-reinforced membrane, where a porous hydrophobic  matrix film is impregnated 
with an ion-conducting ionomer. As a support material for the  matrix, ceramic films
of metal oxides \cite{tejedor-2005,vichi-1999-ceramic-membrane}, 
silica xerogels  \cite{colomer-2003}, 
surface-modified silica films \cite{iacob-2016-ionome-in-nanopore}, 
polypropylene,  polytetrafluoroethylene (PTFE), polycarbonate,  polysulfone, and microglass
fiber fleece (see Refs.~\cite{liu-2003,rodgers-2008} and references therein)  have been used. 
Matrix reinforcement improves the mechanical stability of the 
modified PEM against water flooding, and increases its resistance to the crossover
of fuel molecules. 
However, the pore walls of the matrix  restrict  macromolecular motion of the ionomer,
 causing a lack of connectivity of the hydrophilic pathways. As a result, 
 the ion diffusion 
{\need{ 
rate
}}
   in  matrix reinforced PEMs is smaller than the diffusion in 
the corresponding bulk ionomers. A simple increase in 
the pore size will not solve this issue, because 
the increased connectivity of the ionomer will  also increase 
  membrane permeability to the feeding gases. 


\begin{figure} [!ht]
\begin{center}
\includegraphics*[width=.99\textwidth]{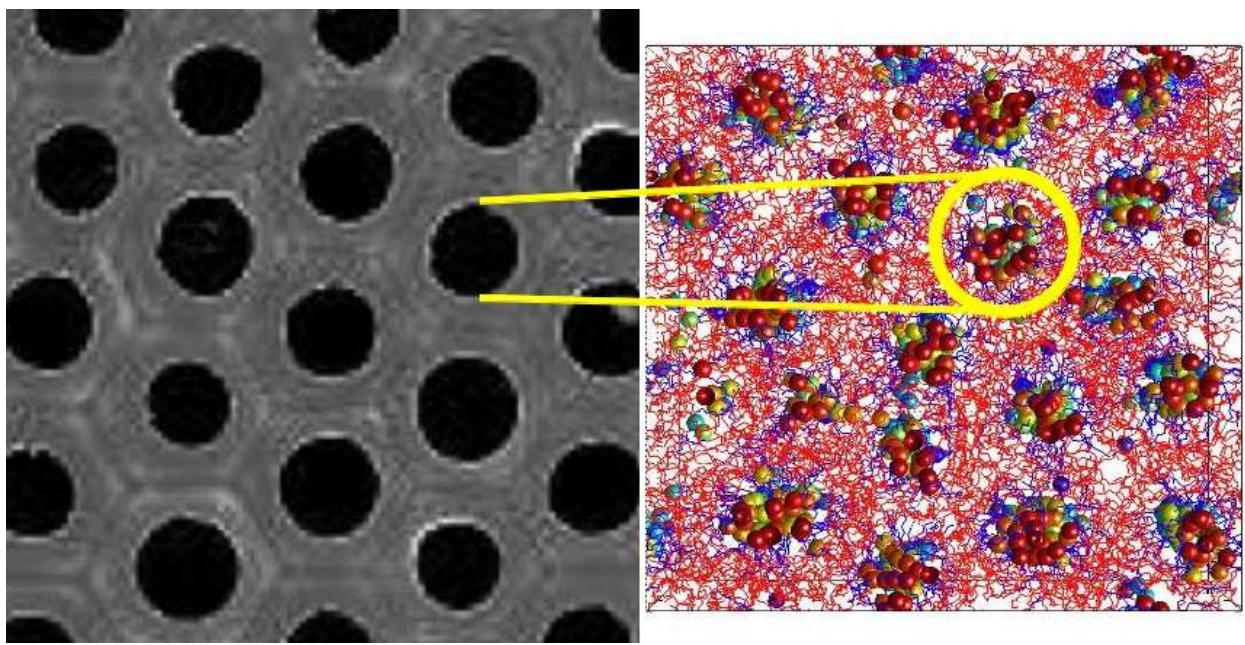}
\end{center}
\vspace{-0.2cm} 
\caption{ 
(Color online) A schematic picture explaining the concept of 
a {\it regular porous matrix reinforced membranes}. 
The right-side colored figure is a top-view snapshot picture from  
Ref.~\cite{allahyarov-2010-poling-pre} and shows the formation of cylindrical 
sulfonate aggregates along the applied electrical field. 
Red lines in the snapshot  correspond to the hydrophobic PTFE backbone of the ionomer, 
and  blue lines correspond to the pendant sidechains. 
Spherical beads denote the positions of sulfonate groups with coloring 
commensurate with their altitude $x$ in the simulation box. 
The left-side picture represents a hard-wall  
 porous matrix with regularly ordered and parallel cylindrical pores.
\label{fig-1-new}
}
\end{figure}

{\need{ 
 An alternative approach  for the matrix reinforced ionomer might be a 
concept of a regular porous matrix with the pore walls grafted 
with pendant sidechains, 
as  schematically illustrated in Figure~\ref{fig-1-new}.
Here the left-side picture represents a hard-wall  
 porous matrix with regularly ordered and parallel cylindrical pores, and  
the right-side picture represents  a top-view snapshot of 
the ordered macrophase separation in a poled Nafion$\textsuperscript{\textregistered}$-like ionomer 
(perfluorinated polymer from DuPont)
from our previous 
work \cite{allahyarov-2010-poling-pre}.  
The ionic diffusion in such poled membranes takes place along the 
cylindrical sulfonate aggregates 
\cite{allahyarov-2010-poling-pre,allahyarov-poling-2009,allahyarov-2011-poling-of-dry,allahyarov-2011-poling-diff-archit}. 
The formation of similar parallel and inverted-micelle like cylinders 
in a Nafion ionomer has been 
reported  in Ref.~\cite{shmidt2007}. This finding, however, was later disputed   
in Refs.~\cite{eikerling-2008,kreuer-2013-water-freezing}     
where an alternative slit geometry for the sulfonate aggregates 
was suggested.
}}

{\need{ 
 In narrow pores the confinement effect of the walls suppresses the freezing point of water   
with respect to pure water 
\cite{cappadonia-1994,cappadonia-1995,rennie-1977-water-freezing,jiao-2009-frozen-water,plazanet-2014-water-freezing-in-nafion,thompson-2006}.  As a result, the  water absorbed in the membrane tends to be in a liquid state at sub-zero temperatures.
In ionomer membranes at low water contents $\lambda$$\le$4.8 the 
confined water   bonds tightly to the sulfonates and becomes strongly structured. This additionally 
prevents the water from freezing at low temperatures
\cite{wan-2014-water-freeze-cold-start-fuel-cell}. In such pores 
the proton transport at sub-zero temperatures probably occurs according to the Grotthuss mechanism. 
}}

\begin{figure} [!ht]
\begin{center}
\includegraphics*[width=.69\textwidth]{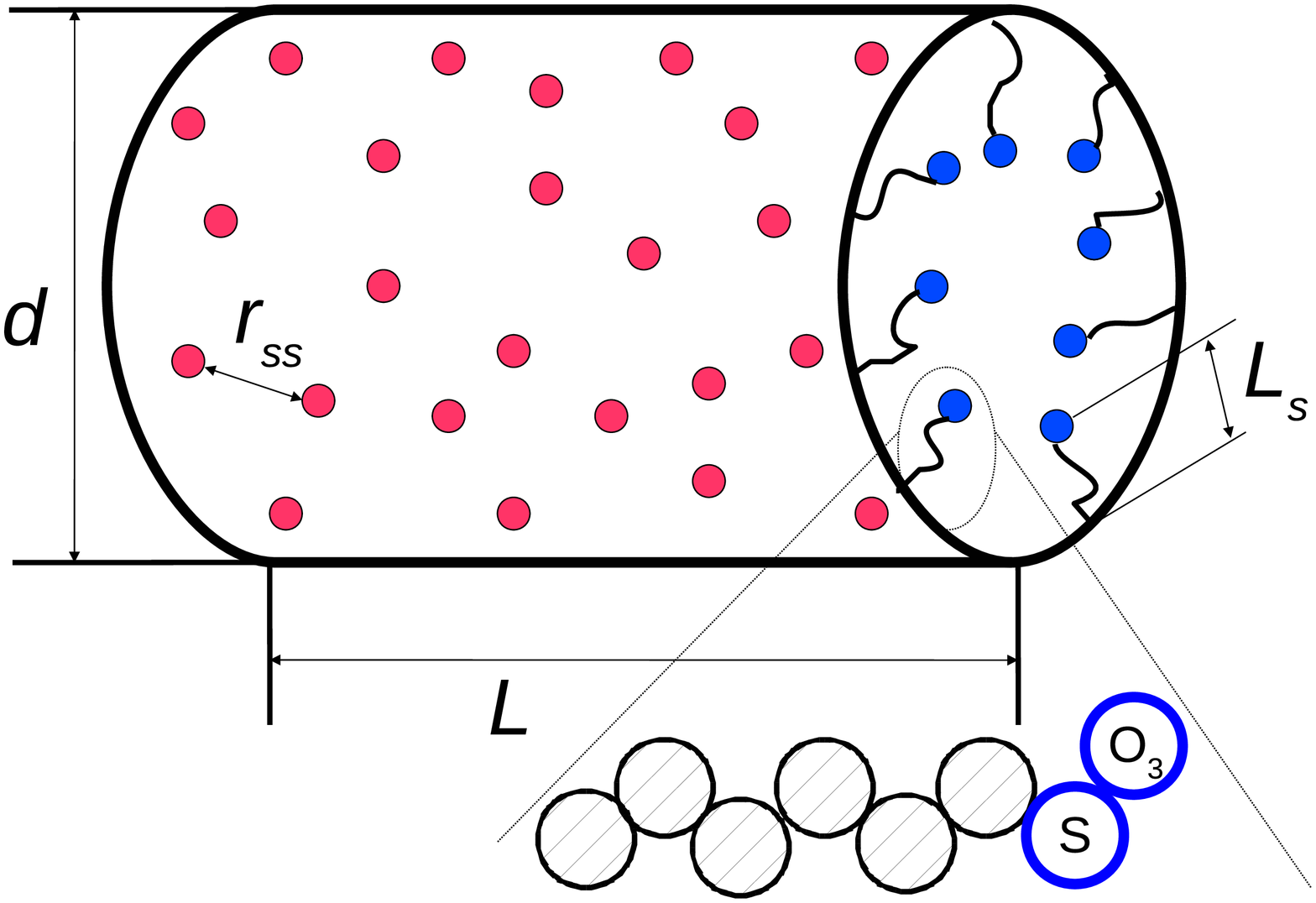}
\end{center}
\vspace{-0.5cm} 
 \caption{ 
(Color online) A schematic representation of a cylindrical  pore 
of a width  $d$ and length $L$, grafted with sidechains
 of length $L_s$ monomers. 
 The anchoring centers of the  sidechains are randomly distributed 
 on the cylindrical surface with an average separation distance $r_{ss}$ 
between the neighboring centers. The hydrophobic parts 
of the sidechain are shown as hatched circles, and 
the hydrophilic terminal groups $S$ (a sulphur atom) 
and O$_3$ (the oxygen group) are drawn in blue. Other details 
are given in the text. 
\label{fig-2-new}
}
\end{figure}

{\need{ 
In this work we consider a single pore with hydrophobic and stiff walls
 which represents the pores in the matrix reinforced membrane 
shown in Figure~\ref{fig-1-new}. 
The pore size is assumed to be in the range of 2-3.7 nm, 
and the pore walls are grafted with terminal groups to provide  similar 
sulfonate aggregates  as in the snapshot picture in  Figure~\ref{fig-1-new}. 
The proposed single pore, in addition to being totally resistant to swelling, 
water flooding, and gas crossover problems, has two additional  remarkable features.
}}
 First,  no external electric field is needed to generate continuous lanes of sulfonates.
A continuous cluster  network  is formed self-consistently through  the sulfonate-sulfonate aggregation   
 supplemented  by the hydrophobic wall-hydrophilic sulfonate and 
wall-water repulsive interactions  \cite{allahyarov-2015,kumar-2005}.
 Second, the water content can be regulated by the pore size, the grafting density and 
 the number of sidechain monomers. These parameters set up a non-homogeneous distribution of the ions 
and water molecules in the pore  \cite{allahyarov-2015,sulbaran-2005,jonsson-1990}. 

Our objective is to analyze systematically  the role of the sidechain monomer number
 $L_s$, the pore diameter $d$, and the water content $\lambda$ (the number of water molecules per terminal group)
 in getting enhanced ion diffusion along the pore axis 
{\need{
compared to the ion diffusion in a bulk Nafion-like ionomer at the same hydration levels.
}} 
We suggest that the flexibility of the sidechains  and the dipolar nature 
of the sidechain sulfonates will greatly assist the ion transport along the pore 
axis, provided that the pore center is free from ion and sulfonate clusters. 
We report high ion diffusion coefficients  for water contents  
$\lambda$$\approx$8--10 and propose a  new scaling rule for the pore 
parameters which guarantees maximal ion diffusion along the pore axis. 
 According to our findings, the sulfonate group protrusion length into the pore center,
 measured as twice the gyration radius of the sidechain with the Flory exponent 1/4, 
should be about 1/3 of the pore radius for securing maximal diffusion rates for the ions.

The rest of the paper is organized as follows. 
In section~2 
we describe the main parameters of
the accepted model for the  pore geometry and anchored sidechains.
The details of the  simulation runs are given in  section~3. 
In section~4 
we consider three different sets of pore parameters for the detailed analysis of
the ion diffusion in charged pores, and discuss the subsequent results. Our conclusions are 
presented in section~5. 








{\mod{ 
\section{Simulation model }
}}
\label{section-model}

A schematic picture describing  our simulation  model for the pore geometry  
 is shown in Figure~\ref{fig-2-new}.
The inner surface of the cylindrical pore of length $L$ and diameter $d$
is randomly grafted with  $N_s$  anchoring centers. 
These centers serve as the attachment points to the sidechains  
 modeled as a spring-bead polymer chain of $L_s$ monomers. 
This number, $L_s$, 
will be referred to hereafter as the sidechain protrusion length.
 A typical structure of a sidechain with $L_s$=8 monomers is
  shown at the bottom of Figure~\ref{fig-2-new}. The shaded sidechain
 particles are hydrophobic and  electrostatically neutral with 
zero charge, 
    while the hydrophilic terminal group monomers at the sidechain 
tip,  inscribed with the letters S for the sulphur atom 
and  O$_3$ for the oxygen group atoms,  
are charged. 
{\need{
We assume for them partial charges $q_S$=+1.1$\vert e \vert$, and 
 $q_{O_3}$=--2.1$\vert e \vert$, 
\cite{allahyarov-2011-poling-of-dry,allahyarov-swelling-2009}  
where $e$ is the electron charge. 
The assigned charges $q_S$ and $q_{O_3}$ depend on the fractional charge distribution along the sidechain. 
For the latter there are different force-field approaches 
\cite{spohr-2004,jang2004,jinnouchi-2003,li-2001-partial-charges,urata-2005,vishnyakov2001} which   
  put the charge $q$=--$\vert e \vert$, necessary to compensate the hydronium charge $q_H$=+$\vert e \vert$,
 either entirely on the  terminal group SO$_3$, or smear it over the sidechain 
monomers. For example,  the following terminal group charges have been considered in the literature: 
 $q_S$=+0.932$\vert e \vert$ and $q_{O_3}$=--1.757$\vert e \vert$ \cite{urata-2005},
 $q_S$=+1.7$\vert e \vert$ and $q_{O_3}$=--2.25$\vert e \vert$ \cite{vishnyakov2001},
 $q_S$=+2.08$\vert e \vert$ and $q_{O_3}$=--2.57$\vert e \vert$ \cite{li-2001-partial-charges},
 $q_S$=+1.284$\vert e \vert$ and $q_{O_3}$=--1.862$\vert e \vert$ \cite{jinnouchi-2003}, 
or $q_S$=+1.0817$\vert e \vert$ and $q_{O_3}$=--1.852$\vert e \vert$ \cite{jang2004}. 
All of these do not fully compensate the  hydronium charge.  
The residual charge $\Delta q$=$q_S + q_{O_3} -\vert e \vert$ 
 is thus distributed between the other sidechain monomers. 
In Ref.~\cite{spohr-2004} with  $q_S$=+1.19$\vert e \vert$ and $q_{O_3}$=--2.19$\vert e \vert$,  
$\Delta q$=0, and thus  the hydronium charge is fully compensated by the charge of the terminal groups. 
In this sense, our choice for  $q_S$  and $q_{O_3}$ is close to the terminal group charges   adopted in Ref.~\cite{spohr-2004}. 
}}

In our model the total negative charge of the $N_s$ terminal groups SO$_3^-$ 
is compensated by the  positive charge of the $N_s$ hydronium ions  modeled as spherical blobs of diameter $\sigma$.
The pore also holds $\lambda N_s$ water molecules H$_2$O which hydrate the terminal groups and liberate the ions. 
In bulk Nafion-like ionomers the parameter $\lambda$  is varied between 3 and 25.

\begin{figure}  [!ht]
\begin{center}
\includegraphics*[width=0.74\textwidth,height=0.4\textwidth]{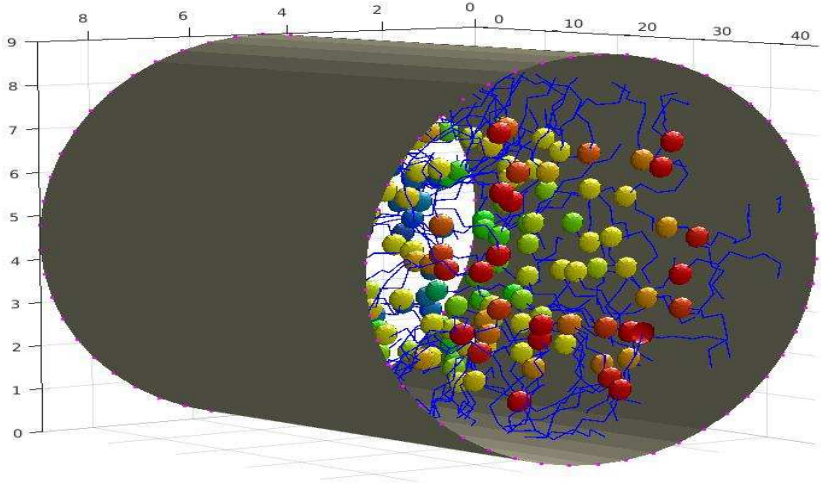}
\end{center}
\caption{(Color online) 
 A typical snapshot picture from the simulation box for the 
 run B4 from Table~\ref{table-abc}. 
Sidechains are drawn in blue. The coloring of  sulfonates  
is commensurate with their distance from the right-side end of the pore. 
 The size of all structural elements is schematic rather than space filling. The distances are
  given in units of the bead diameter $\sigma$=0.35 nm.   
 \label{fig-3-new}}
\end{figure}

The pore geometry parameters $L$ and $d$, and the number of grafting points $N_s$ 
define the grafting density of the sidechains $n_s$=$N_s/(\pi d L)$,
 which can be also referred to as the surface charge density of the pore. 
The average neighbor-to-neighbor distance between 
the anchoring points is
\begin{equation}
r_{ss} \approx \frac{1}{\sqrt{n_s}}
\label{r-anchor}
\end{equation}
For short protrusion lengths with $L_s$=2 monomers, 
when terminal groups are in the vicinity  of the pore walls, 
the parameter $r_{ss}$ essentially has the meaning of the  average distance $a_1$ 
between the neighboring sulfonate groups. For Nafion-like ionomers
both these quantities usually take values  from 
0.6 to 1.2 nm. 
However, for long sidechains with 
$L_s$$\ge$3 monomers, when the sulfonate groups tend to protrude  into the  pore volume 
and form compact clusters, $a_1$ is defined from the position of 
the first maximum of the pair distribution function of 
sulfonates $g_{OO}(r)$ (here $O$ represents terminal group oxygens O$_3$). 
{\need{
In general, the smaller $a_1$ is, the more robust is  
proton  hopping among neighboring sulfonate groups via intermediate water molecules. 
Considering that the minimal value of $a_1$ may not be smaller than 0.6 nm because of the electrostatic repulsion 
between negative SO$_3^{-}$ groups,  at least one water molecule should be present between neighboring sulfonate
groups. This should enable accepting a proton from one sulfonate and passing it to another sulfonate
 \cite{tsampas-2006,perrin-2007,choi2005,eikerling-2001,marx-2006,spohr-2002}. 
}}

{\mod{ 
\section{Simulation Details}
}}
 \label{section-details}
 
We employ a coarse-grained approach  for the sidechain
in the framework of the united-atom representation for the  CF$_{2}$
 groups (from the blob number 1 to the blob number $L_s$-2 of the sidechain) 
\cite{vishnyakov2001,yamamoto,allahyarov-2011-diff-archit}, 
and  for the sulphur   atom S and  the oxygen group O$_3$
of sulfonates  \cite{allahyarov-2011-diff-archit,allahyarov-dry-ionomer-2007}. 
 All united-atom groups are modeled as Lennard-Jones (LJ) monomers 
 with a diameter $\sigma$=0.35 nm and 
 6-12 LJ interactions among them. 
The sidechain constituents  are additionally subjected to stretching, bending  and dihedral
forces. The force-field details  are given in our previous work 
\cite{allahyarov-swelling-2009,allahyarov-2011-diff-archit,allahyarov-stretching-2009} and 
 agree in most instances with the Nafion sidechain 
model of Paddison \cite{Paddison1}.  A  brief description of the force field includes four components.
First, the total potential energy of the sidechain polymers  is 
\begin{equation}
U(\vec r) = \displaystyle\sum_{i}{U_b^i} +
\displaystyle\sum_{j}{U_{\theta}^j} + 
\displaystyle\sum_{m}{U_{\varphi}^m} + 
\displaystyle\sum_{k,l}{U_{nb}{\left( |\vec r_k -\vec r_l| \right)}} 
\label{eq1} 
\end{equation}
where $(\vec r_1, \vec r_2,...,\vec r_N)$ are the
three-dimensional position vectors of the $N$ particles in the system, 
the index $i$   in the two-body bond-stretching potential $U_b^i$ runs over all bonds, 
the index $j$   in the three-body angle-bending potential $U_{\theta}^j$ 
 runs over all bond angles, 
the index $m$  in the four-body dihedral component of the interaction energy runs
 over all torsional angles, 
and indices $k$, $l$ in the non-bonded (Lennard-Jones and Coulomb) 
potential run over all  force-center pairs in the system.
{\need{ 
 With the sidechain equilibrium bond length $b$=0.44$\sigma$ and the equilibrium bond angle 
$\theta$=110$^o$, the fully stretched sidechain with a dihedral angle $\alpha$=0 has 
a physical length $l_s$=$L_s b \sin(\theta/2)$. For the maximal number of sidechain monomers 
$L_s$=18 we get $l_s$=2.27 nm. 
}}  
Second, the water is modeled as a TIP3P liquid  
\cite{allahyarov-2011-diff-archit,tip5p-tip3p}, which 
has explicit charges  $q_H$ = +0.417$\vert e \vert$ on the hydrogen atom, 
and $q_O$ = -2$q_H$  on the oxygen atom. 
The  distance between hydrogen and oxygen atoms is  $r_{OH}$=0.0957 nm,
 and the angle between OH bonds
is $\theta_{HOH}$=104.52$^o$. The TIP3P model does not allow for inclusion of the 
non-classical Grotthuss transport of protons 
\cite{eikerling-2008,kornyshev-2003,kornyshev-2009-theory-md}. The ion transport 
in our simulation model represent a classical {\it en-masse} diffusion of ions 
which is affected by the dynamical rearrangement of flexible sidechains \cite{thompson-2006}.    
Third, standard Lorentz-Berthelot combining rules are used in the LJ interactions
between monomers $i$ and $j$ such that the interaction potential between units $i$ and $j$  has a minimum at 
the  separation distance $r_{ij}=2^{1/6} \sigma_{ij}$ where  $\sigma_{ij}$ is defined as 
  $\sigma_{ij} = \left( \sigma_{ii} + \sigma_{jj} \right)$/2,  and the depth of the interaction at this minimum 
is $\epsilon_{ij} = \sqrt{ \epsilon_{ii} \epsilon_{jj}}$. 
For hydrophobic monomers $\epsilon_{ii}$=0.2 kcal/mol, and for hydrophilic 
monomers    $\epsilon_{jj}$=0.1 kcal/mol coefficients were used in the LJ interaction 
potentials. 

The numerical part of our study consists of three series of
 simulation runs. In the  first series of runs,  a pore with grafted sidechains is created 
and the system is subsequently  equilibrated during  runs of 1 ns  
 at  constant volume and  temperature. In these runs the
charges of the ions, water atoms and terminal groups were set to zero, and 
the equilibration was  done following the hydrophobic and hydrophilic 
interactions in the system. 
In the second series of runs, the charges were turned on,
and the system was equilibrated for another 1 ns of simulation time 
under a constant pressure condition $P$=1 atm. The details of this procedure 
are explained in section~4. 
Finally, in the third series of runs, 
 the necessary statistics on the sidechain conformations and the ion 
diffusion were gathered during the final  runs of 50 ns duration. 

During  the production runs, the simulated system was
 coupled  to a Langevin thermostat with a friction coefficient
$\gamma=2 $ ps$^{-1}$ and a Gaussian white-noise force of strength
$6k_{B}T\gamma$.  The equations of motion were
 integrated using the velocity Verlet algorithm with a time step of
 0.5 fs. One dimensional periodic  boundary conditions were 
 imposed on the system along the pore long axis, which coincides with the 
$x$-axis of the system. The  translational 
replicas of the fundamental cell pore correspond to an infinite cylindrical 
pore. 
The long-range electrostatic interactions between charged particles are handled using a one-dimensional 
Lekner-like summation
 \cite{paul-paddison-2005-pore,montoro-abascal-1998-1D-ewalds,mazar,gronbech-jensen-1d-lekner-1997}. 
{\need{ 
For simplicity, it is assumed that there is no dielectric discontinuity between the interior of the pore 
 and the wall material. Otherwise, the interaction of the ions with their 
image charges will add more complexity to our simulations. 
 }}

The wall-monomer interaction is treated using the particle-micropore interaction potential 
given by Tjatjopoulos {\it et al.} in Ref.~\cite{ion-pore-LJ-interaction}, 
further developed in Ref.~\cite{tang-2001}, and described in detail in our previous 
work \cite{allahyarov-2015}. The wall is considered smooth with  no atomic structure on it
\cite{kim-2006}. 
 The wall-ionomer interaction is hydrophobic for the hydrophobic segments of the  sidechain.  
The wall-terminal group, wall-ion, and wall-water molecule   interactions  are hydrophilic.

For the dielectric constant  of the pore interior we use a distance-dependent 
effective permittivity function 
$
\epsilon(r) = 
 1 + 
\left( 
         \epsilon_B - 1  
\right)
\left(
    1 - r/\sigma 
\right)^{10} / 
\left(
    1 + (r/\sigma)^{10}  
\right)
$
similar to Refs.~\cite{allahyarov-2015,allahyarov-2011-diff-archit}. 
This function gradually increases from $\epsilon=1$ at small monomer-monomer separations 
to the bulk ionomer value $\epsilon_B$=4 
{\need{ 
for $r/\sigma \gg$1.
}}
{\need{ 
 Although the best approximation for $\epsilon(r)$ can be a matter of debate 
\cite{paddison-2003-review,paul-paddison-2005-pore,taylor-1992}, it
is clear that the permittivity must increase with distance 
to  account for the polarization effects of the neutral and hydrophobic parts of the sidechain 
 when the interaction between the ions is calculated. 
The suggested $\epsilon(r)$ does not  overestimate the Coulomb interactions
in the system when explicit water exists in the pore. All  water confinement, polarization, and screening effects are explicitly taken into 
account in our set-up through the explicit ion-water and water-water Coulomb interactions. 
}}

{\mod{ 
\section{Simulation Results}
}}
\label{sim-results}

In total, there are five  system parameters, namely 
 the sidechain protrusion length  $L_s$ monomers, 
 the pore diameter $d$, 
 the water content  $\lambda$, 
 the pore length $L$, 
 and 
 the number of grafting points   $N_s$ on the pore surface,
which fully define the pore structure.  
The fixing of the last two parameters, the 
pore length to $L$=46$\sigma$ ($L$=16 nm) and the 
number of grafting sites to $N_s$=200, allows us to reduce the 
number of system variables from five to three. 
Subsequently,    
we will analyze the  following three separate cases.  
The {\bf case A} with ($\overline{L_s},d,\lambda$), 
the {\bf case B} with ($L_s,\overline{d},\lambda$), and 
the {\bf case C} with ($L_s,d,\overline{\lambda}$), where
the overlined variables will be kept constant while  the two other 
parameters inside the parentheses will be treated as running variables.
The set-up parameters for the cases {\bf A}, {\bf B}, and {\bf C}
 are given in Table~\ref{table-abc}. 

{\need{
The pore parameters  $d$ and  $\lambda$ listed in Table~\ref{table-abc} are coupled 
to each other by the constant pressure condition $P$=1.0 atm. For  the 
one-component system in restricted geometries the pressure can 
be calculated using the virial theorem \cite{kumar-2005,kumar-2007}. However, 
this method is not efficient for multi-component systems like confined and hydrated ionomers. 
 In these complex systems  the water mostly forms hydration shells around sulfonate groups and hydronium ions 
where its density 
is about  10-20 percent higher  than the bulk water \cite{medasani-2014}. 
As a result of this, the local pressure  deviates strongly from the average pressure in the system \cite{chialvo-1999}, 
and thus the virial pressure method becomes less accurate.  
}}

{\need{ 
We calculate the pressure $P$ from the cumulative force  $F_{wall}$ of the 
hydronium-wall and water-wall interactions. For the {\bf cases A} and {\bf C}
the pressure $P$=$ F_{wall}/(\pi d L)$=1 atm was achieved by treating 
the pore size $d$ as a running parameter.  
For the {\bf case B} all runs were started with the minimal water content $\lambda$=1. 
This parameter subsequently was increased through adding more water molecules until 
 $P$=1 atm is reached in the pore. 
}}

{\need{
It should be noted that the calculated water content $\lambda$ in the
{\bf case B}, and  the pore size $d$ in  the {\bf cases A} and {\bf C} 
under the condition $P$=1 atm might be different from 
experimental realizations  because of the non-polarizable water model used 
in our simulations. The precise polarizable water models \cite{walbran-2001}  
might result in different water densities in the solvation shells of ions and terminal groups. 
This  is still an open question that deserves close attention, 
though it is out of the scope of our current model. 
}}

 The {\bf case A}, where the dependence of the 
ion diffusion on the pore size at various water contents will
be elaborated, closely  mimics the  pore swelling  process in ionomer membranes.
In the {\bf case B} the principal role of the sidechain protrusion length $L_s$ 
will be revealed in pores with a fixed size $d$. In this case the change of $L_s$ 
is complemented by  the change in the water content $\lambda$. Lastly, 
the {\bf case C} is devoted to the role of the pore size $d$ in ion diffusion. 
For each of these cases we will determine optimal pore parameters 
for obtaining  maximal ion diffusion coefficients $D$  and $D_x$, which are defined below. 

A snapshot picture taken  from the simulation cell 
for run B4 is shown in Figure~\ref{fig-3-new}. Here, 
along with the sidechains drawn in blue and the 
sulfonate groups SO$_3^-$ shown as spheres,  
 the pore also accommodates 
$N_s$ monovalent ions and $\lambda N_s$ water molecules, which are not shown 
 in this figure. 
Here two different types of effect take place. 
The sidechain-specific effects are  their strong electrostatic fields, which
polarize the water and attract  
{\need{ 
 hydronium H$_3$O$^+$ ions 
}}
 to the sulfonates, 
and their affinity to  clustering due to the dipolar nature of  SO$_3^-$ groups  and mutual
SO$_3^-$--SO$_3^-$ attraction mediated by the ions.  
The pore-specific effects are the pore wall hydrophobicity and the pore size which regulate
the distribution of terminal groups, water molecules  and ions. 

\begin{table}[!ht]
\begin{center}
\caption{\label{table-abc} Simulation parameters for the 
{\bf case A}, {\bf case B}, and {\bf case C} runs. 
All distances are given in nm units.
 The pore length is $L$=16 nm and the number of grafting points on 
the pore surface is $N_s$=200. 
The parameters $d^*$ and $r_{ss}^*$ for the {\bf case A}
 correspond to the reduced value of $N_s=100$. 
{\need{ 
$L_s$ is the number of monomer units per sidechain. The physical length 
$l_s$ of a stretched  sidechain with zero dihedral angle is $l_s$=$L_s b \sin(\theta/2)$.
}}
Overlined quantities on the left column represent the fixed pore parameter for the corresponding runs. 
}
\begin{tabular}[t]{lccccccc}
\hline 
\hline 
case A runs            & A1  & A2   &  A3 &  A4 & A5 &   A6 & A7  \\ 
\hline 
{\bf $ \overline L_s $}  & {\bf 4 }  & {\bf 4}    & {\bf 4}   & {\bf 4}   &{\bf 4}    & {\bf 4}   & {\bf 4}  \\
$d$  
 & 2.03 & 2.33 & 2.70     &2.89   & 3.12 & 3.37 & 3.55  \\
 $ \lambda $  & 1   & 3    & 6   & 8   & 10    & 13   & 15  \\
$r_{ss}$  & 0.65 & 0.71 & 0.77  & 0.80   & 0.84  & 0.87 & 0.90   \\
$d^*$  
   & 1.54 & 1.75 & 2.01 & 2.17   & 2.30 & 2.49 & 2.61  \\
$r_{ss}^*$ 
    & 0.78 & 0.84  & 0.92  & 0.96   & 0.99  & 1.04 & 1.07   \\
$ $  &     &       &       &        &       &      &  \\
\hline 
case B runs             & B1  & B2   &  B3 & B4 & B5 & B6 & B7  \\   
\hline 
 $L_s$       & 2   & 4 & 6   & 8    & 10  & 12  & 16  \\
{\bf $ \overline d$}  & {\bf 2.89} & {\bf 2.89} & {\bf 2.89}     &{\bf 2.89}   & {\bf 2.89} & {\bf 2.89} & {\bf 2.89}  \\
$\lambda $  & 9.3 & 8 & 6.7 & 5.7  & 4.2 & 2.8 & 1   \\
$r_{ss}$  & 0.80 & 0.80 & 0.80  & 0.80   & 0.80  & 0.80 & 0.80   \\
$ $  &     &       &       &        &       &      &  \\
\hline 
case C runs            & C1  & C2   &  C3 & C4 & C5 & C6 & C7  \\ 
\hline 
$L_s$         & 4    & 8    & 10   & 12  & 14  & 16   & 18  \\
$d$       & 2.89   & 3.15 & 3.30   & 3.40  & 3.50  & 3.60   & 3.70   \\
{\bf $ \overline \lambda $ } & {\bf 8} & {\bf 8} & {\bf 8} & {\bf 8 } & {\bf 8} & {\bf 8} &  {\bf 8 }  \\
$r_{ss}$   & 0.80    & 0.84  & 0.86  & 0.88 & 0.89 & 0.91  & 0.92  \\
\hline 
\hline 
\end{tabular}
\end{center}
\end{table}

 For each simulation run listed in Table~\ref{table-abc}, 
the ion  self-diffusion coefficients $D$  and $D_x$  were calculated using 
 their mean square of displacements, 
\begin{eqnarray}
6 D t   = \frac{1}{N_s} \displaystyle\sum_{i=1}^{N_s} \langle  |\vec r_i(t) - \vec r_i(0)|^2 \rangle \nonumber \\
2 D_x t = \frac{1}{N_s} \displaystyle\sum_{i=1}^{N_s}  \langle |x_i(t) - x_i(0)|^2 〉\rangle 
\label{diffusion}
\end{eqnarray}
where $r_i(t)$ and $x_i(t)$ are the 3D and 1D-axial positions of 
the atom $i$ at the simulation time $t$. 
Eq.(\ref{diffusion}) defines only the {\it en-masse} (vehicular) ion diffusion. 
The structural proton diffusion through  Grotthuss hopping,
which is not included in our simulation model, but which  can  be modeled using
  empirical valence bond (EVB) based water models \cite{spohr-2002,walbran-2001,evb-model},
would increase the total diffusion rate of ions.

It should be noted that the pore diameters considered in 
Table~\ref{table-abc}  
are typical for Nafion-like ionomers, and are   
larger than the size $\approx$0.4--1.1 nm of narrow carbon nanotubes.  
Thus, for the run parameters listed in Table~\ref{table-abc}
  we do not enter the regime of single-file diffusion 
\cite{mukherjee-2010, striolo-2007,striolo-2006}. 
Moreover,  the presence of the ions and grafted chains with hydrophilic ends 
 is a suppressing factor to any occasional movement 
of hydrogen-bonded water clusters along the pore axis 
\cite{joseph-aluru-2008,dellago-hummer-2003}. 

{\mod{ 
\subsection{Case A simulations for  ($\overline L_s$,$d$,$\lambda$)}
}}

The runs A1-A7 from Table~\ref{table-abc} are consistent with the ionomer pore swelling 
when excess  water is absorbed by the PEM membrane. That is why we consider shorter
sidechains with $L_s$=4 monomers which correspond to the sidechain protrusion length about 3--4 \AA \, in 
the pores of  Nafion-like membranes 
\cite{paddison-2003-review}.

\begin{figure}  [!ht]
\begin{center}
\includegraphics*[width=0.49\textwidth]{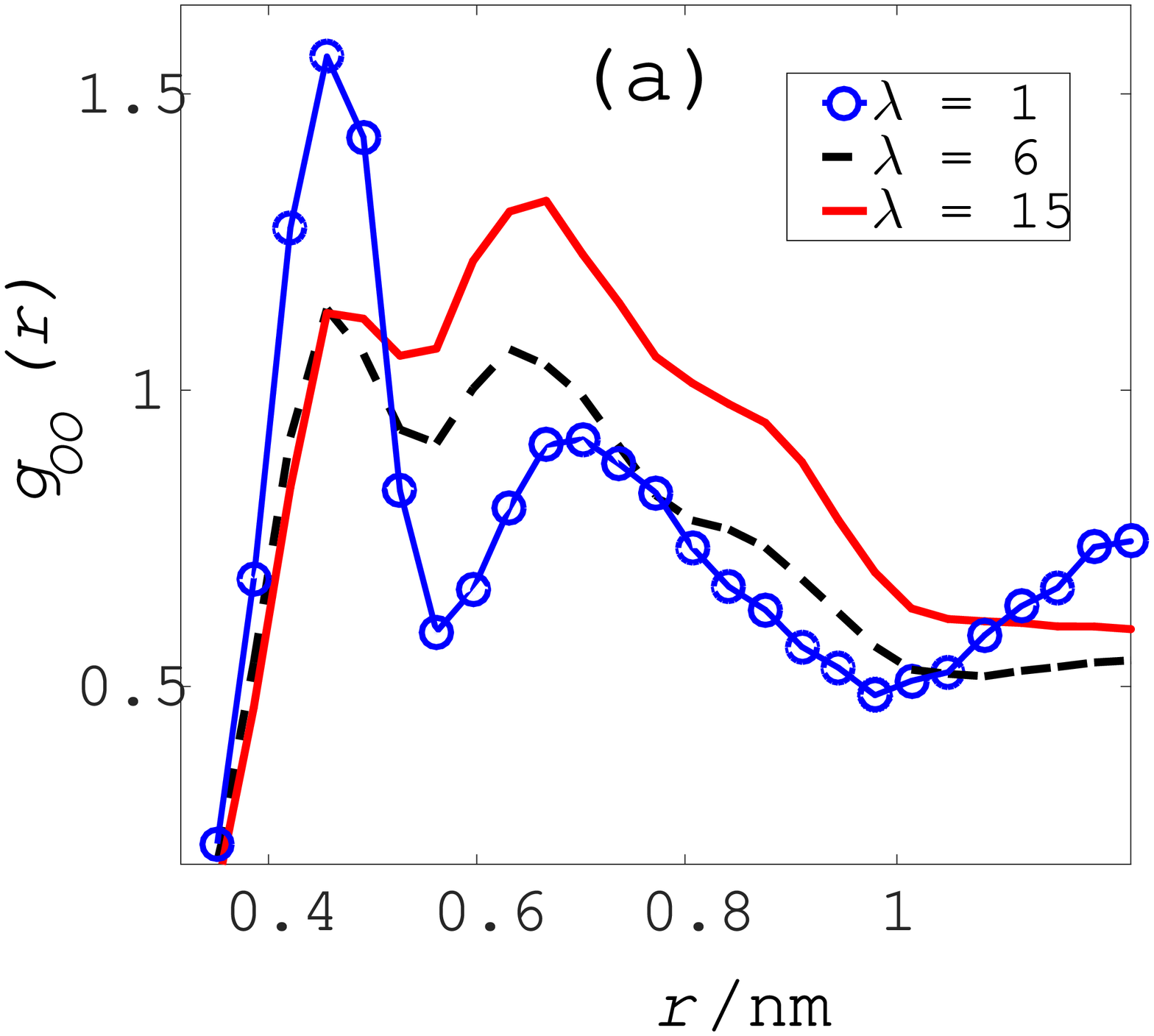} 
\includegraphics*[width=0.49\textwidth]{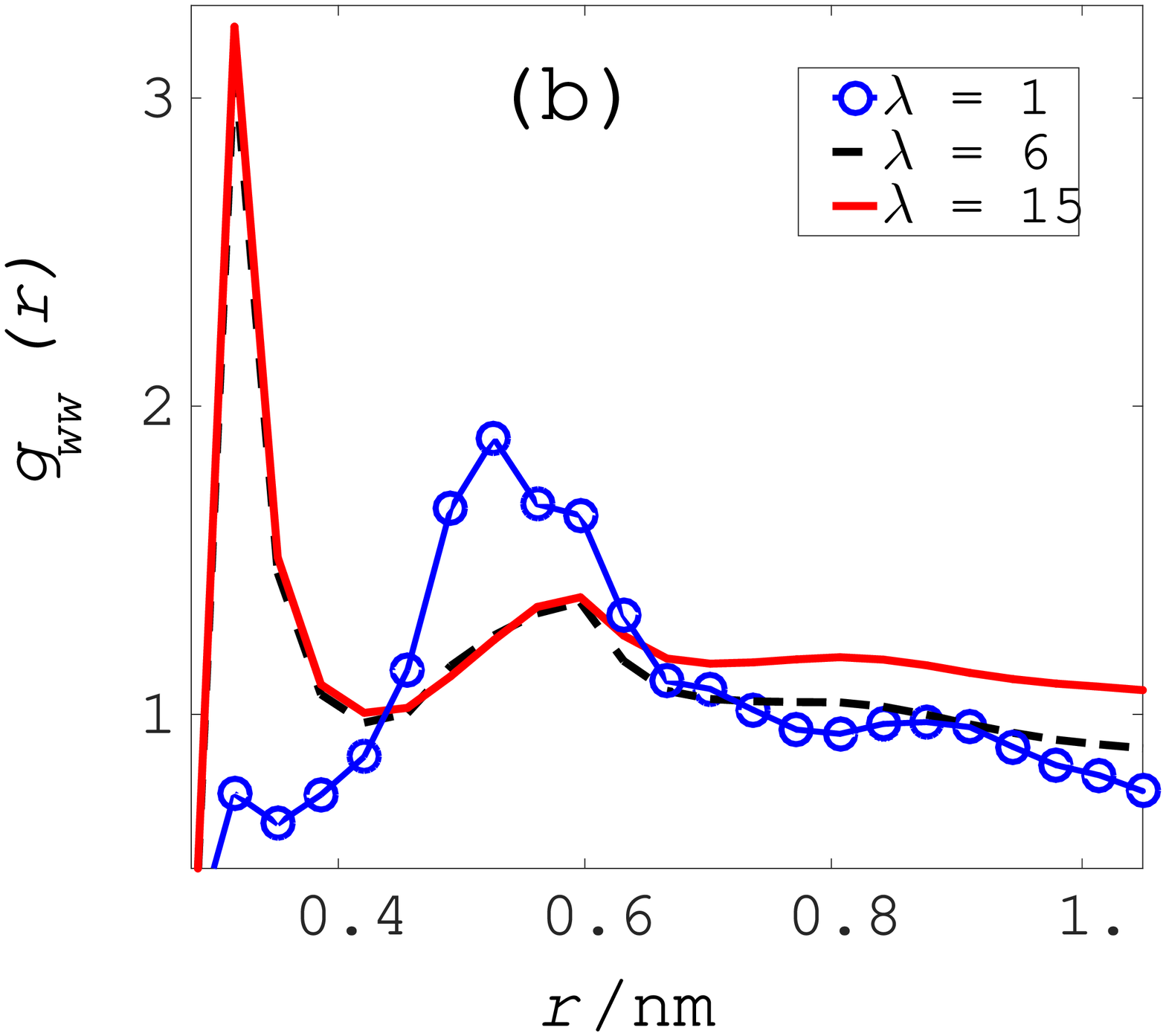}
\end{center}
\vspace{-0.5cm}
\caption{(Color online)  Simulation results 
for the sulfonate-sulfonate $g_{OO}(r)$ (a) and water-water $g_{ww}(r)$ (b) 
 pair distribution functions  for the runs 
A1 ($\lambda$=1), A3 ($\lambda$=6), and A7 ($\lambda$=15) from 
 Table~\ref{table-abc}.
Other parameters: $L_s$=4 monomers and $N_s$=200. 
\label{fig-4-new}}
\end{figure}

We first plot the  pair distribution functions $g_{OO}(r)$ of the terminal group oxygens 
in Figure~\ref{fig-4-new}a. The   pair distribution functions are defined as
\begin{equation}
g_{jj} = 
\frac{1}{4 \pi r^2 \Delta r N_s}
\displaystyle\sum_{i=1}^{N_s} \delta{\left(\vec r - \vec r_i^{(j)} \right)} 
\,\, \theta \left( \vert \vec r -  \vec r_i^{(j)} \vert  - \Delta r\right)
\label{grr}
\end{equation}
where $\vec r_i^{(j)}$ denotes the position of the particle $i$ of the type $j$, 
 $j$=O for the oxygen groups of sulfonates,  
$\delta(r)$ is the Dirac function, and $\theta(r)$ is the Heaviside step function. 
{\need{  
The two maxima of the $g_{OO}(r)$  at separation distances $a_1$$\approx$0.45 nm
and $a_2$$\approx$0.7 nm correspond to the first and 
second nearest neighbor sulfonate groups, respectively.
The neighboring sulfonate groups share their hydronium ions in the sulfonate clusters of size 2 nm. 
In swollen pores  $g_{OO}(a_1)$ decreases whereas $g_{OO}(a_2)$ increases, 
which is different from the case of the bulk ionomer, 
 where both these quantities decrease at high solvation levels 
 \cite{allahyarov-2010-poling-pre,blake2005}.
This behavior is a result of the increase in the sidechain anchoring distance $r_{ss}$ 
which makes it harder for the terminal groups to come closer and share a hydronium ion at a 
separation distance $a_1$. As a consequence, in swollen pores the 
sidechain tips mostly share their counterions  at the separation distance $a_2$.
For larger $\lambda$ the  discrepancy between $r_{ss}$  and $a_2$ increases, which indicates 
that the sidechains stretch to form sulfonate clusters. 
}}
 
The water-water pair distribution functions $g_{ww}(r)$, defined by  
the Eq.(\ref{grr}) with $j$=$w$ for the water oxygens,
  are shown in Figure~\ref{fig-4-new}b. 
For the low water contents $\lambda$$\leq$3 the most expected 
water-water separation distance is  about 0.5 nm, which,  
 more likely, characterizes the separation distance 
between water molecules that share the same ion or  sulfonate group. 
At high water contents $\lambda$, hydration shells   are formed  
around the charged entities together with the water clusters being formed in the pore center. 
Both these effects appear as the formation of a  huge maximum 
at the touching water-water separation distance $a_1$= 0.35 nm for $\lambda$$\geq$6 
 in Figure~\ref{fig-4-new}b.
 The position of the second maximum of $g_{ww}(r)$ at 
$a_2$$\approx$0.6 nm is related to the radius of the second solvation layer  
in water clusters.   

\begin{figure}  [!ht]
\begin{center}
\includegraphics*[width=0.49\textwidth,height=0.4\textwidth]{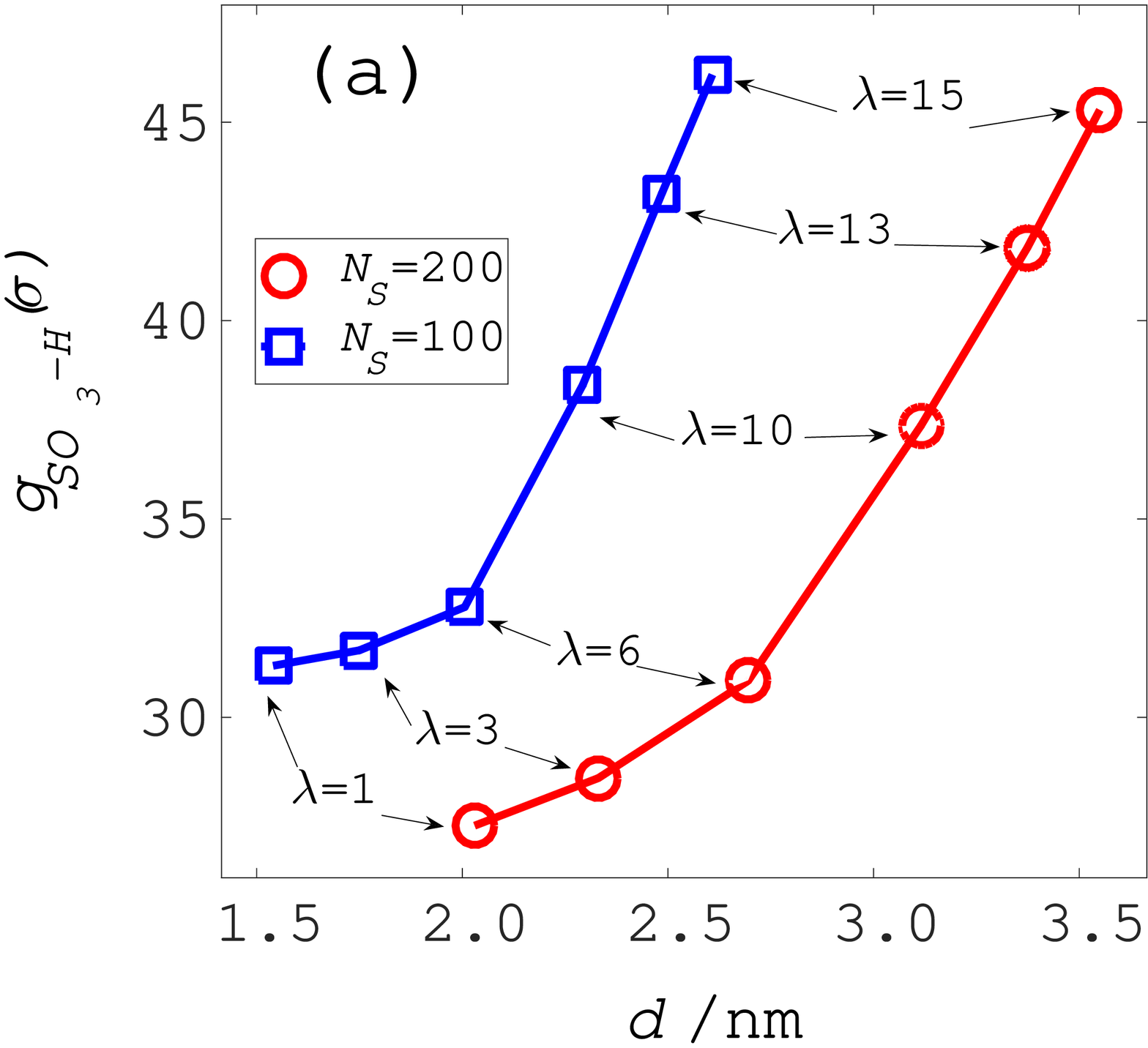}
\includegraphics*[width=0.49\textwidth,height=0.4\textwidth]{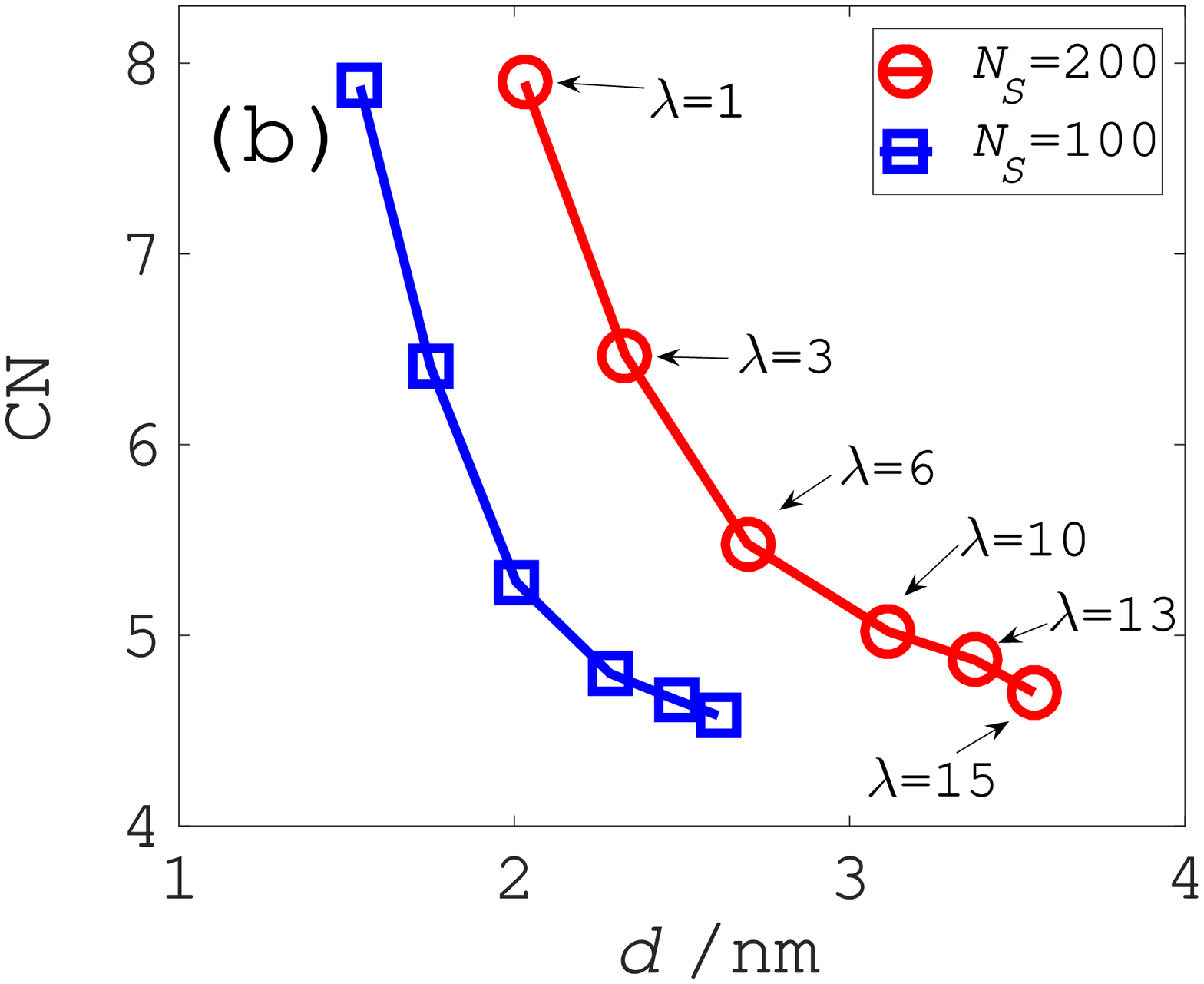}
\end{center}
\vspace{-0.75cm}
\caption{(Color online) (a) Simulation results for the ion-sulfonate association
parameter  $g_{SO_3-H}(\sigma)$ for the runs A1-A7 from 
Table~\ref{table-abc}. 
{\need{ 
(b) Coordination number CN for sulfonates, see Eq.(\ref{CN}) in the text, 
for the runs A1-A7 from Table~\ref{table-abc}. 
}}
The number of grafted sidechains is $N_S$=100 for the blue line with squares, 
and $N_S$=200 for the red line with circles. 
\label{fig-5-new}} 
\end{figure}

{\need{ 
The ion-sulfonate association parameter, measured as the height of  the SO$_3^-$-ion 
pair distribution function $g_{SO_3-H}(r)$ at $r$=$\sigma$, 
is plotted in  Figure~\ref{fig-5-new}a. This parameter holds 
information on  
how strongly the electrostatic field of  sulfonate groups  disturbs the distribution of hydronium ions in the pore. 
We see that such disturbance increases in swollen pores.   
This tendency, discussed in Refs.~\cite{eikerling-kornyshev-2001,eikerling-2001}, 
 can be justified  by the enhancement of the sulfonate-ion interaction 
\begin{equation}
U_i(r)= -\frac{e^2}{\epsilon_s r} e^{-\frac{r}{\, \, \,\, r_{D}}}
\label{db}
\end{equation}
in wide pores. 
In Eq.(\ref{db})  $\epsilon_s$ is the permittivity of the medium and includes  polarization effects of  water 
in  swollen pores, $r_{D}$ is the Debye screening length for the sulfonate groups  and hydronium ions,  
\begin{equation}
r_D=\sqrt{\frac{k_BT \epsilon_0 \epsilon_s}{ n_e e^2}}=
\sqrt{\frac{k_BT \epsilon_0\epsilon_s d}{8 n_s e^2}} \approx
r_{ss} \sqrt{\frac{k_BT \epsilon_0  \epsilon_s d}{ 8 e^2}} \approx
0.008 r_{ss} \sqrt{ \frac{\epsilon_s d}{\sigma}}
\label{r-debye}
\end{equation}
where $\epsilon_0$ is the permittivity of free space, $n_e$ is the concentration of all charges (SO$_3^-$ and H$_s$O$^+$) in the system 
and is defined as $n_e$=$2 N_s/(\pi (d/2)^2 L)$=$8 N_s/( \pi d^2 L)$=$8 n_s/d$. 
Using Eq.(\ref{r-debye})  we get 
$r_{D1}$=0.028 nm in  the unswollen     pore of run A1, and 
$r_{D2}$=0.115 nm in  the fully swollen pore of run A7 
by assuming that  $\epsilon_s$=5 in the first, and $\epsilon_s$=25 in the second cases.   
Then, for the relative change of the association strength $U_2(\sigma)/U_1(\sigma)$  we get $U_2/U_1$$\gg$1
which is a clear indication of the fact that 
in swollen pores the distribution of hydroniums is strongly disturbed by the electric field of terminal groups. 
This is exactly what is observed in Figure~\ref{fig-5-new}a, which, however, should not be interpreted as the  
localization effect of H$_3$O$^+$ ions around the sulfonate groups. \\
The localization or delocalization effects of the
ions around their host terminal groups can be evaluated through their  coordination number (CN) 
\begin{equation}
{\mbox{CN}}= \rho_H^0 \int_{\sigma}^{R_m} g_{SO_3-H}(r) \, 4 \pi r^2 \, dr
\label{CN}
\end{equation}
shown in Figure~\ref{fig-5-new}b. Here  
 $\rho_H^0$=$N_S/V$ is the average sulfonate density in the pore of volume  $V=\pi d^2 L/4$, 
and $R_m$ is the position of the first minimum of $g_{SO_3-H}(r)$. The decrease of CN
  in swollen pores, as seen from Figure~\ref{fig-5-new}b,  corresponds to 
the expected ion delocalization effect due to the water screening effects.
}}

{\need{
Figure~\ref{fig-5-new} also shows the weakening of the ion-sulfonate association  $g_{SO_3-H}(\sigma)$
 when the pore surface charge density  is increased. 
This is seen, for example,  for $d$=2.3 nm, where
 the  $g_{SO_3-H}(\sigma)$  is about 38 at $\lambda$=10 and  $N_s$=100,
compared to its  value       about 28 at $\lambda$=3  and  $N_s$=200. 
This ion-sulfonate association weakening  can be explained,  again,  using the 
screening arguments given in Eq.(\ref{r-debye}). In the pores with higher $N_s$ and  lower 
$\lambda$, the parameters  $r_{ss}$ and $\epsilon_s$, 
and therefore the screening length $r_D$,  become smaller. 
 At reduced $r_D$ the charges of all ions are strongly screened, and 
thus the opposite charges in the pore are loosely associated. 
As a result of this, the CN of the sulfonate groups also decreases in pores with lower
surface charge densities, as seen in Figure~\ref{fig-5-new}b. 
%
%
In total, 
the competition between the increased influence of the sulfonate groups on the 
distribution of the hydronium ions and the increased delocalization of the hydronium ions 
in swollen pores determines the optimal pore diameter when the hydronium diffusion 
along the pore axis reaches its  maximum value. 
}}

\begin{figure}  [!ht]
\begin{center}
\includegraphics*[width=0.79\textwidth,height=0.59\textwidth]{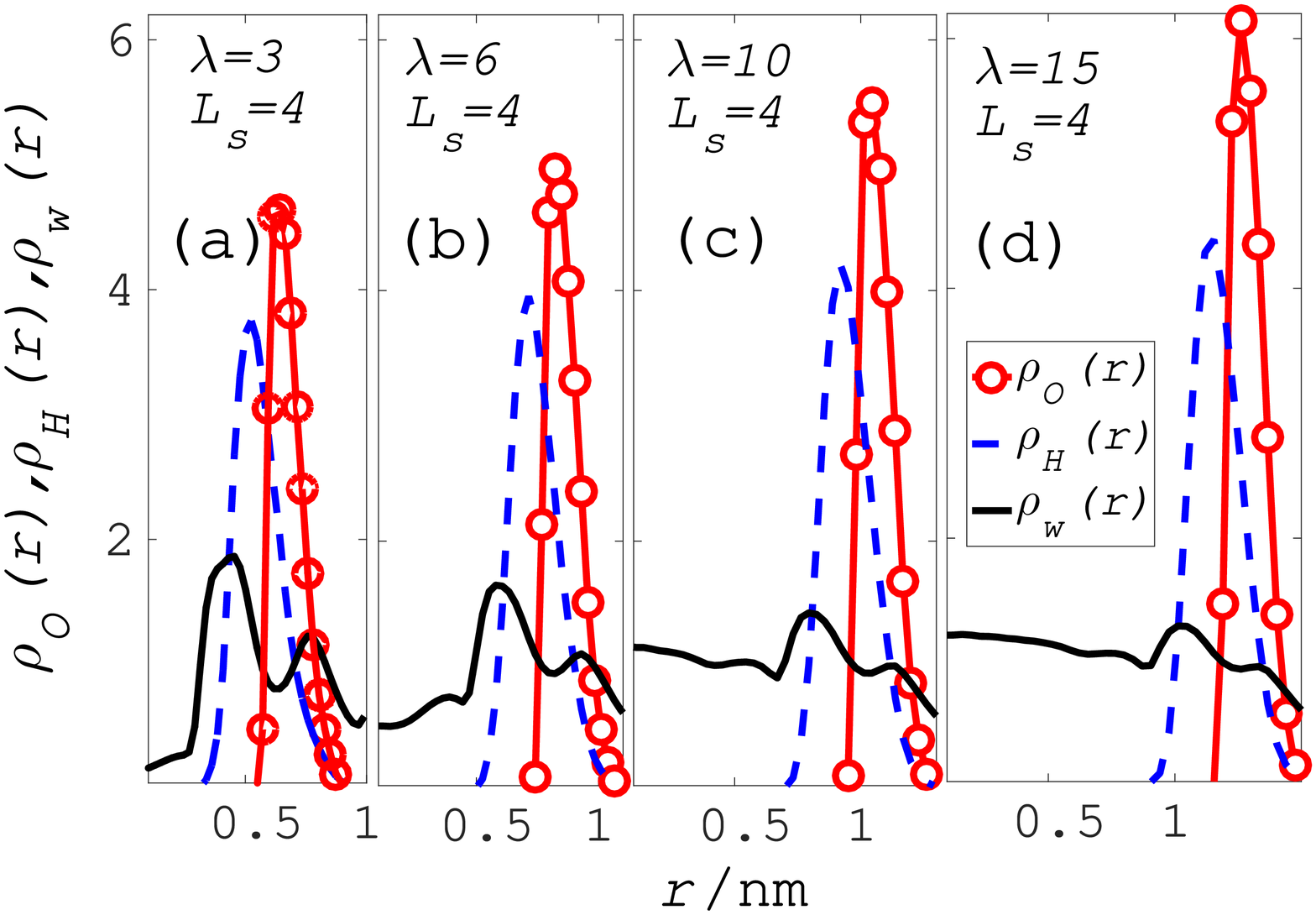}
\end{center}
\vspace{-0.7cm}
\caption{(Color online)  Normalized radial distribution
  functions $\rho_j(r)/\rho_j^0$ for the 
{\need{ 
hydronium 
}}
 ions ($j$=$H$), water molecules ($j$=$w$) and 
  the sulfonate oxygens ($j$=$O$), 
as a function of the radial distance $r$ from  the pore center
 for the simulation runs A2 (a), A3 (b), A5 (c), and A7 (d) from
  Table~\ref{table-abc}.   The normalizing
  factor $\rho_j^0=N_j/V$, where 
$N_j$ is the  number of particles of the sort $i$. 
Full red lines with circles- sulfonate oxygens, dashed blue lines- ions, 
solid black lines- water molecules.
 \label{fig-6-new}}
\end{figure}

The radial distribution functions, defined as  
\begin{equation}
\rho_j(r^{(j)}) =
\frac{1}{2 \pi r \Delta r L  N_s}
\displaystyle\sum_{i=1}^{N_s} \delta{\left( r - r_i^{(j)} \right)} 
\,\, \theta \left( \vert  r -   r_i^{(j)} \vert  - \Delta r\right)
\label{rad-func}
\end{equation}
are shown in Figure~\ref{fig-6-new} for  
the sulfonate oxygens ($j$=$O$), 
{\need{ 
 hydronium 
}}
ions ($j$=$H$), and the water molecules ($j$=$w$). 
In Eq.(\ref{rad-func}) $r$ is the radial distance from the pore axis
 ($r$=0 corresponds to the pore center), 
and $r_i^{(j)}$ is the radial distance of the particle $i$ of the sort $j$. 
In Figure~\ref{fig-6-new} the radial distributions are normalized by the average density of 
particles $j$ in the pore. 
The ion profiles $\rho_H(r)$ appears to be completely restricted to 
the pore boundary regions and no free ions penetrate to the pore center. 
This result differs from the ion radial distributions calculated  
using approaches based on  Poisson-Boltzmann theory  in  
charged pores \cite{eikerling-kornyshev-2001,yang-pintauro-2004,pintauro-1995}. In the latter 
the ion profile is non-zero in the whole pore volume. 
{\need{ 
The exclusion of ions from the pore center in  Figure~\ref{fig-6-new} effectively means that 
in charged pores with shorter grafted  sidechains,  the 
Grotthuss  mechanism of structural proton transfer is restricted to the
pore boundary areas. 
}}

The radial distribution of water molecules $\rho_w(r)$ 
shown in Figure~\ref{fig-6-new} displays strong water structuring effects in the pore.
 First, in narrow pores with $\lambda$$\leq$6, the water  
 mostly accumulates in the hydration shells of ions and sulfonates 
where it is strongly polarized. This is a consequence of the high
 surface-to-volume ratio effect in narrow pores, which works against 
the water clustering in the pore center.
 In the larger pores the water shows bulk-like properties in the pore central area. 

The increase in the height of the maxima in 
$\rho_O(r)$ in Figure~\ref{fig-6-new} in larger pores 
 is related to the frustration of sidechains. 
When $r_{ss}$ becomes larger in swollen pores, 
the sidechains start to stretch out so that  their tips 
can  share free ions.   
Also, in larger pores the sidechain-pore wall interaction becomes stronger:  
while the hydrophobic part of the sidechain becomes  more attracted to the pore wall, 
the hydrophilic tip segment of the sidechain becomes more repelled from the  pore wall.  
An increased frustration of sidechains in larger pores, together with the 
enhanced sulfonate-ion association effect shown in Figure~\ref{fig-5-new}a, brings the 
ions to the vicinity of the pore walls. This is seen from the increase 
in the height of the 
maxima in  $\rho_H(r)$   in Figure~\ref{fig-6-new}. 
As a result of these frustration-localization effects the 
flexibility of sidechains in larger pores degrades.   

\begin{figure}  [!ht]
\begin{center}
\includegraphics*[width=0.5\textwidth]{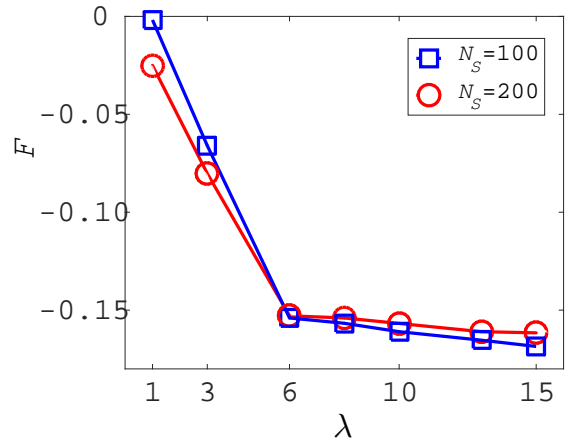}
\end{center}
\vspace{-0.8cm}
\caption{(Color online) The sulfonate group SO$_3^-$  dipole orientation
  parameter $F$ as a function of the water content $\lambda$ for 
the runs A1-A7 from Table~\ref{table-abc}. 
The number of grafted sidechains is $N_S$=100 for the blue line with squares, 
and $N_S$=200 for the red line with circles.
\label{fig-7-new}}
\end{figure}

The orientation of the dipole moment of the  SO$_3^-$ head group, 
characterized by the function  
\begin{equation}
   F = \frac{3 \langle \cos^2(\theta) \rangle -1}{2} \,\,\, ,
\label{dipole}
\end{equation}
is shown in Figure~\ref{fig-7-new}, as a function of the water content $\lambda$. 
In Eq.(\ref{dipole})  $\theta$ is the angle  between the  SO$_3^-$  dipole moment and 
 the long axis {\it x}  of the pore, and $F$=--0.5 is expected for 
a limiting case when  all the SO$_3^-$ dipoles point towards the pore center. 
The orientation parameter $F$ shows a steep decrease when $\lambda$ is increased  from     
1 to 6, which is followed by a weak change in $F(\lambda)$ 
for $\lambda$$\ge$6. Such behavior can be explained  by  
assuming that at $\lambda$=6 all ions develop a compact hydration shell
 which  is strongly repelled from the walls of the larger pores. 
This results in the reorientation of the sulfonate dipoles towards the pore 
center. On the other hand, the accumulation of bulk water in the pore center
has only a negligible effect on the sulfonate dipole reorientation. 

\begin{figure} [!ht]
\begin{center}
\includegraphics*[width=0.69\textwidth]{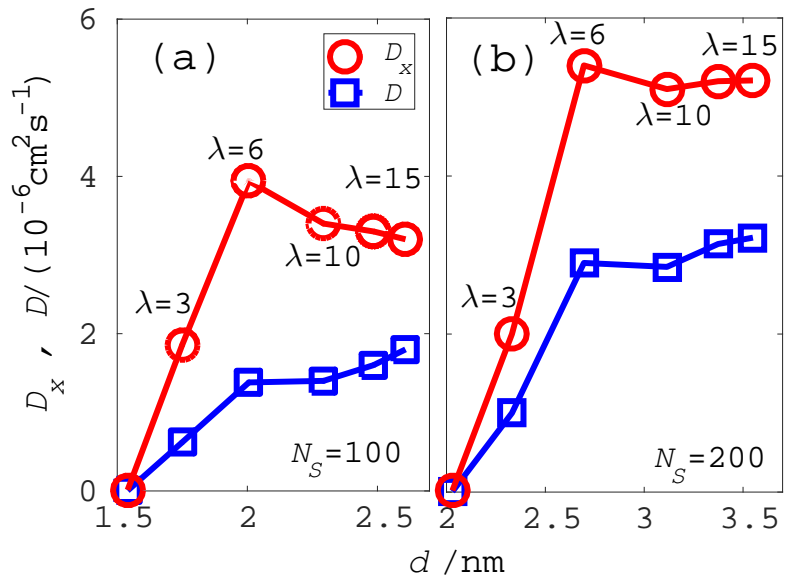}
\end{center}
\caption{(Color online)
 The total ion  diffusion coefficient $D$ 
(blue line with squares), and the 1D ion diffusion coefficient $D_x$ along the pore longitudinal axis $x$ 
(red lines with circles) for  the runs A1-A7 from Table~\ref{table-abc}. 
The number of grafted sidechains is $N_s$=100 in (a) 
and $N_s$=200 in (b). 
 \label{fig-8-new}}
\end{figure}

The ion self-diffusion coefficients  $D$  for  3D ion diffusion in the pore, and $D_x$  
for 1D ion diffusion along the pore axis $x$,
defined by Eq.(\ref{diffusion}),  are   shown in Figure~\ref{fig-8-new} 
as a function of the pore size $d$. 
For all the hydration levels for the runs A1-A7, simulation results 
show $D_x>D$, showing that  the ion diffusion along the pore axis is higher than the 
total ion diffusion in the pore  \cite{kerisit-2009}.
While the 3D diffusion has  almost a  monotonic  
 dependence on $\lambda$, the 1D diffusion reaches a maximum at $\lambda$=6. 
Such nonlinear behavior of the $D_x(d)$ is a consequence of 
the following two factors:  
First, as the pore swells, the increase in the grafting  parameter
$r_{ss}$ 
and the decrease in the sidechain flexibility constrain 
the ion diffusion along the pore. 
Second, as the pore shrinks,  the low amount of water, 
insufficient  to hydrate the ions, becomes  highly structured 
 and polarized by the charges. As a consequence, the ions 
become immobile in the structured water environment. 
  The maximum of the ion diffusion, hence,  happens in between  these
two extreme too-large and too-narrow
pore cases.  
Figure~\ref{fig-8-new} also reveals that the surface charges of the pores facilitate 
 ion diffusion in the pore. This is because of the electrostatic delocalization 
effect of ions due to the 
 weakening of the ion-sulfonate association at high $n_s$. 
In total,  in the  pores with shorter sidechains
 the  ion diffusion is limited to the pore wall area, and its 
component along the pore axis 
is maximal if the electrostatically  delocalized ion 
possesses  a hydration shell with about six water molecules.

{\need{ 
It is known that at sufficiently longer times the diffusion coefficient of the ions in the radial 
direction of the pore becomes very small, $D_{r}$$\to$0, thus 
the full diffusion coefficient 
\begin{equation}
D = \frac{2}{3} D_r + \frac{1}{3} D_x
\label{asymptotic}
\end{equation}
has an asymptotic limit  $D_x/D$=3. In our simulations this limit has been reached 
in the narrow pores of runs A1 and A2 for $N_s$=100, and in the narrow pore of the run A1 for $N_s$=200.
In  the larger pores of runs A3-A7 the ratio $D_x/D$ decreases from 3 to around 1.6--1.8. Therefore, 
 runs much longer than 50 ns are needed to approach the asymptotic limit  $D_x/D$=3.    
}}

{\mod{ 
\subsection{Case B simulations for ($L_s$,$\overline{d}$,$\lambda$)}
}}
 In this section  we focus on the role of the sidechain protrusion length $L_s$ in 
providing fast ion diffusion along the pore axis.  
The runs B1-B7 from Table~\ref{table-abc} were carried out at the constant 
pore size $d$=2.89 nm. The chain protrusion length $L_s$ was increased  from
  2 monomers to 16 monomers
with a  simultaneous decrease of  the  water content $\lambda$ from about 9 to 1.
The shortest $L_s$=2 monomers 
corresponds to a completely hydrophilic sidechain consisting of  only the 
two charged entities of the terminal group SO$_3^-$.
The tips of the longest sidechain with $L_s$=16 monomers  
can reach the pore center and thus are capable of forming hydrophilic sulfonate clusters  there.  
The results of the previous section indicated that 
 the ions  mostly occupy the  pore wall areas in the case of shorter sidechains $L_s$. 
For the longer $L_s$  it is expected that the  
 ions would reach the pore center by following their host sulfonate groups. 
In both cases the ion diffusion depends on the state of water in the pore.
In other words, a  priori,  it is not evident at which $L_s$ and $\lambda$ the ion 
diffusion will be maximal.

\begin{figure}[!ht]   
\begin{center}
\includegraphics*[width=0.49\textwidth,height=0.41\textwidth]{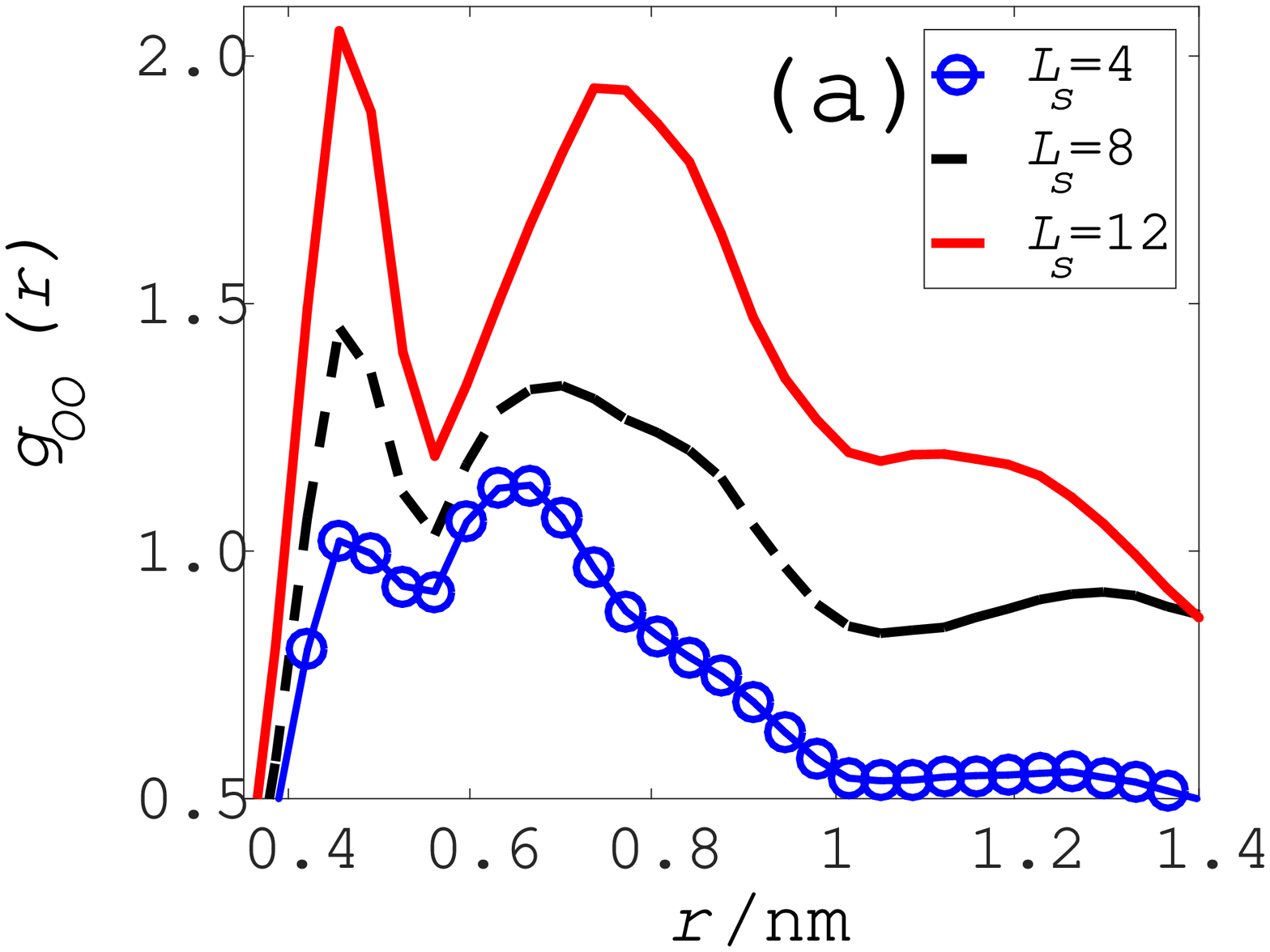} 
\includegraphics*[width=0.49\textwidth,height=0.41\textwidth]{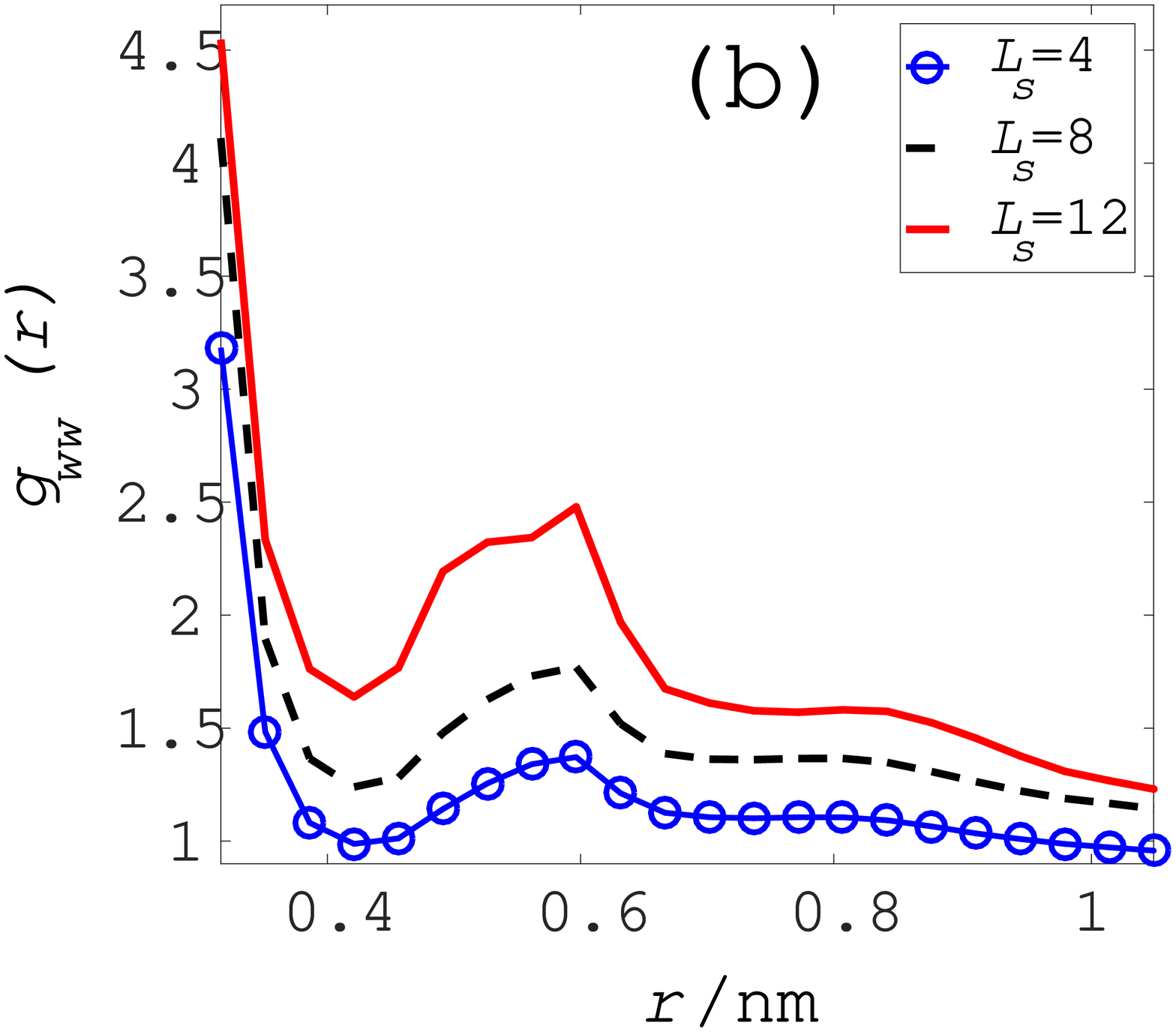}
\end{center}
\vspace{-0.7cm}
\caption{(Color online) Simulation results for the    
sulfonate-sulfonate $g_{OO}(r)$ (a) and 
the water-water $g_{ww}(r)$ (b)
  pair distribution 
functions for the runs B2 ($L_s$=4 monomers), B4 ($L_s$=8 monomers), and B6
 ($L_s$=12 monomers)  from Table~\ref{table-abc}.
\label{fig-9-new}}
\end{figure}

The sulfonate-sulfonate pair distribution functions $g_{00}(r)$
 shown in Figure~\ref{fig-9-new}a 
clearly  indicate an enhancement of the  sulfonate clustering 
 as  the sidechain  protrusion $L_s$ is increased. 
 This result is fully expected, because long sidechains have 
more flexibility to form clusters in the pore volume. As the clusters grow in size, 
 more ions and water molecules are  shared between 
the neighboring sulfonate groups. This explains     
the shift of the position of the second maximum of $g_{00}(r)$, $a_2$, to the larger 
sulfonate-sulfonate separations. 
Since long sidechain  protrusion 
lengths $L_s$ also mean lower values for $\lambda$ in the pore, 
it is obvious that  sulfonate 
clustering features are essentially regulated by the protrusion length $L_s$ 
rather than by the water content $\lambda$.  

\begin{figure}[!ht]
\begin{center}
\includegraphics*[width=0.99\textwidth,height=0.61\textwidth]{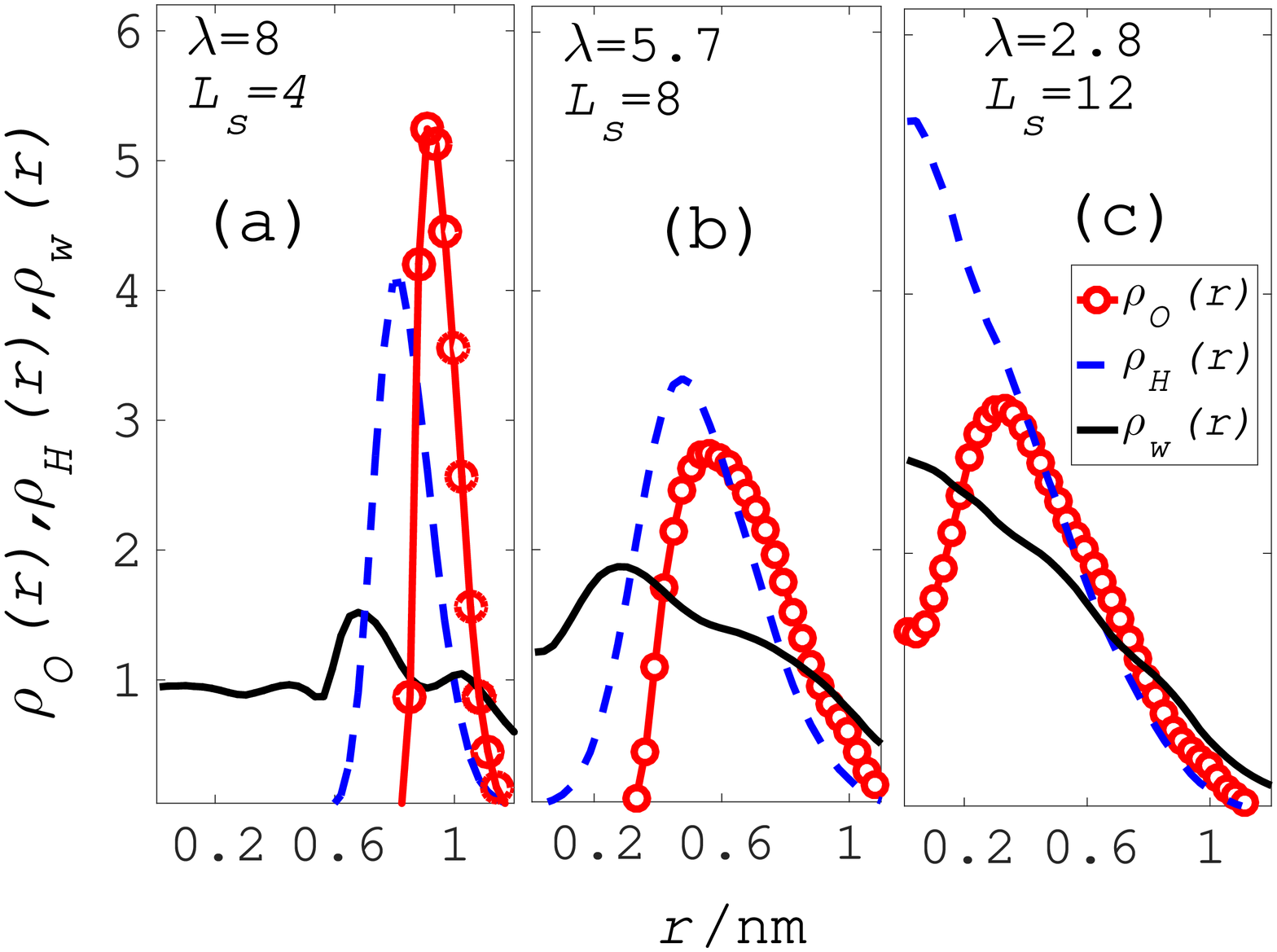}
\end{center}
\vspace{-0.7cm}
\caption{(Color online)   
Normalized radial distribution
  functions $\rho_j(r)/\rho_j^0$ for the 
{\need{ 
 hydronium 
}}
ions  ($j$=$H$), water molecules ($j$=$w$) and 
  the sulfonate oxygens ($j$=$O$), 
as a function of the radial distance $r$ from  the pore center
 for the simulation runs B2 (a), B4 (b), and B6 (c) from
   Table~\ref{table-abc}.   The normalizing
  factor $\rho_j^0=N_j/V$, where $V=\pi d^2 L/4$ is the pore volume,
  and $N_j$ is the  number of particles of the sort $i$. 
Full red lines with circles- sulfonates oxygens, dashed blue lines- ions, 
solid black lines- water molecules.
 \label{fig-10-new}}
\end{figure}

The water pair distribution functions $g_{ww}(r)$ also show higher water cluster 
formations for the long sidechains, as seen in  Figure~\ref{fig-9-new}b.  
Even if the long sidechains  demand less  water to fill the pore,  
it seems that all the available water is collected in the  
pore central area where it  easily  merges into larger clusters. 
This conclusion is supported by  
the radial distribution profiles of ions, water molecules and 
sulfonates, shown in Figure~\ref{fig-10-new}. 
Indeed, as the  protrusion length $L_s$ increases, 
 most of the water molecules cluster in the pore center, 
perhaps by being attracted there by the sulfonate tips of the protruding  
sidechains and the ions. 
For the sidechain monomer number $L_s\leq$8 the radial distributions in 
 Figure~\ref{fig-10-new}a and Figure~\ref{fig-10-new}b 
  look qualitatively  the same as the densities 
shown in Figure~\ref{fig-6-new} for
$\lambda$$\geq$6: the water mostly occupies the pore central area, and the 
ions mostly lie between the water and sulfonate clusters. 
However, for the long chains with the number of monomers   $L_s\geq$12, when the 
sulfonate groups reach the pore center, the ions have their maximal density 
on the $x$-axis. 
 As seen from  Figure~\ref{fig-10-new}c, in this region 
the difference between the radial densities of the sulfonate oxygens $\rho_O(r)$ and 
ions $\rho_H(r)$ gives rise to the charge separation in the pore. 
The central part of the pore  has a positive charge whereas the rest of the pore is charged 
negatively. 
This result is the direct consequence of
the  long sidechains:  a pore with the long sidechains, when the 
chains partly occupy the pore wall area, 
behaves like a narrow pore with shorter sidechains.
{\need{ 
 In narrow pores, when 
their radius is comparable with the sidechain length, 
the counterions enter the pore center 
generating a charge separation between the central  and  wall areas of the pore.
}} 
Their diffusion along the pore axis, however, depends on 
the sulfonate clustering structure in that area.

\begin{figure}[!ht]
\begin{center}
\includegraphics*[width=0.59\textwidth]{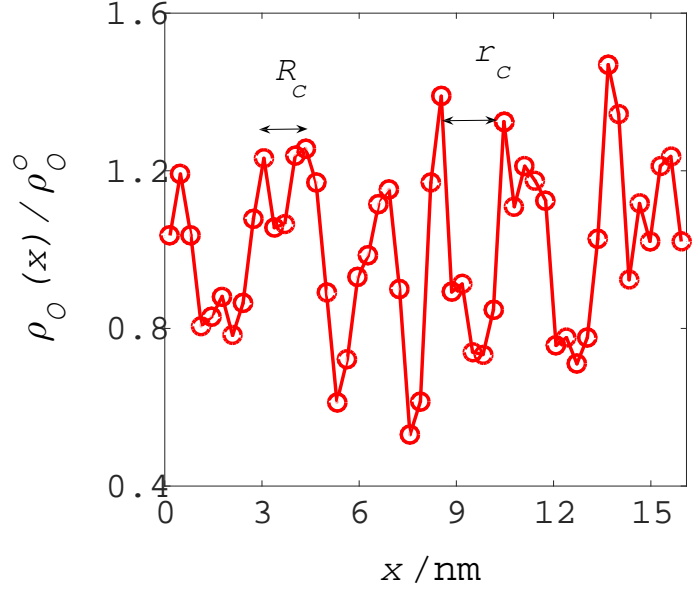}
\end{center}
\vspace{-0.7cm}
\caption{(Color online)  Normalized density of sulfonates 
$\rho_O(x)/\rho_O^0$ along the pore longitudinal axis $x$ for 
the run B6  from 
Table~\ref{table-abc}. The sidechain length is 
  $L_s$=12 monomers. The sulfonates form clusters of the size $R_c$ separated by a distance 
  $r_c$ \label{fig-11-new}
}
\end{figure}

The $x$-axis profile of sulfonate clustering, defined as 
\begin{equation}
\rho_O(x) = 
\frac{1}{\pi d^2 \Delta x }
\displaystyle\sum\limits_{i=1}^{N_s} \delta{\left( x - x_i \right)} 
\,\, \theta \left( \vert  x -   x_i \vert  - \Delta x\right)
\label{rhox}
\end{equation}
where $x_i$ is the $x$-coordinate of sulfonates, 
is shown in Figure~\ref{fig-11-new} for the run B6 with $L_s$=12 monomers
(the case of  radial charge separation in the pore). The sulfonate clusters 
self-organize   into compact formations along the pore axis
with the  average cluster size $r_c$$\approx$2 nm and a  
separation distance between them of  $R_c$=2 nm.  For $N_S$=200 sidechains along the 
$L$=16 nm long pore, a rough estimate gives about $N_c$$\approx$40--50 
 sulfonates per cluster formation.  

\begin{figure}[!ht]
\begin{center}
\includegraphics*[width=0.59\textwidth]{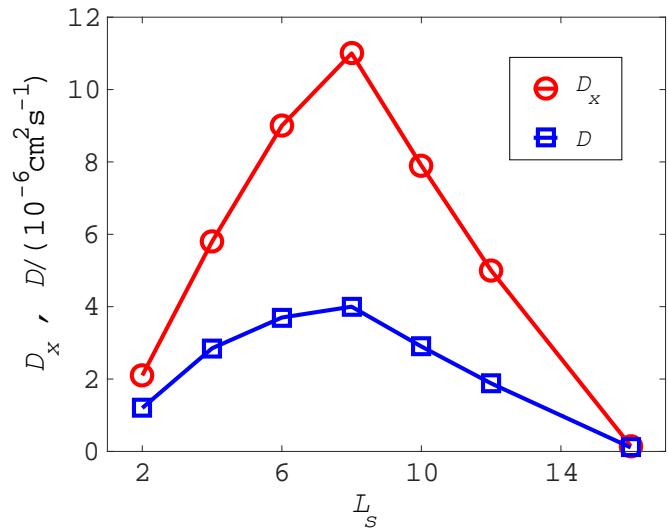}
\end{center}
\vspace{-0.7cm}
\caption{(Color online)   
The total ion  diffusion coefficient $D$ 
(blue line with squares), and the 1D ion diffusion 
coefficient $D_x$ along the pore longitudinal axis $x$ 
(red lines with circles)
for  the runs B1-B7 from  Table~\ref{table-abc}. 
The number of grafted sidechains is  $N_s$=200. 
 \label{fig-12-new}}
\end{figure}

Calculated ion diffusion coefficients $D$ and $D_x$ are shown in  Figure~\ref{fig-12-new}
 for the pore size $d$=2.89 nm and the sidechain anchoring distance $r_{ss}$=0.8 nm.
Both diffusion coefficients  show a nonmonotonic dependence on the sidechain 
 protrusion length $L_s$.
The maximal ion diffusion is observed for the run B4 
with $\lambda$=5.7 and $L_s$=8 monomers. 
According to the findings of Paddison 
 \cite{paddison-2003-review}, who calculated  ion diffusion in the 
framework of non-equilibrium statistical mechanics, 
the ion diffusion should decrease when the 
protrusion length of the anchored sidechains is increased. 
We believe that the difference between our findings  and the results of
 Ref.~\cite{paddison-2003-review} is based on the fact that we 
explicitly treat the dynamics and clustering behavior of the water, ions, and the sulfonates,
which are not included in the statistical mechanical theory of
Ref.~\cite{paddison-2003-review}. 
The fact that the ion diffusion for the run B7 
with $L_s$=16 monomers is very small compared to its value 
for the run B4 with $L_s$=8 monomers, implies that the clustering of sulfonates
 and ions in the pore center 
is a degrading factor for the free ion diffusing in the pore. 
Therefore, for getting faster ion diffusion, it seems that the 
central pore area should be free from ions and sulfonates. 
At the same time, similar to the {\bf case A} runs in the previous section, 
the ions should be partly delocalized from their host sulfonates and 
moderately hydrated.


{\mod{ 
\subsection{Case C simulations for ($L_s$,$d$,$\overline{\lambda}$)}
}}

\begin{figure}[!ht]
\begin{center}
\includegraphics*[width=0.79\textwidth]{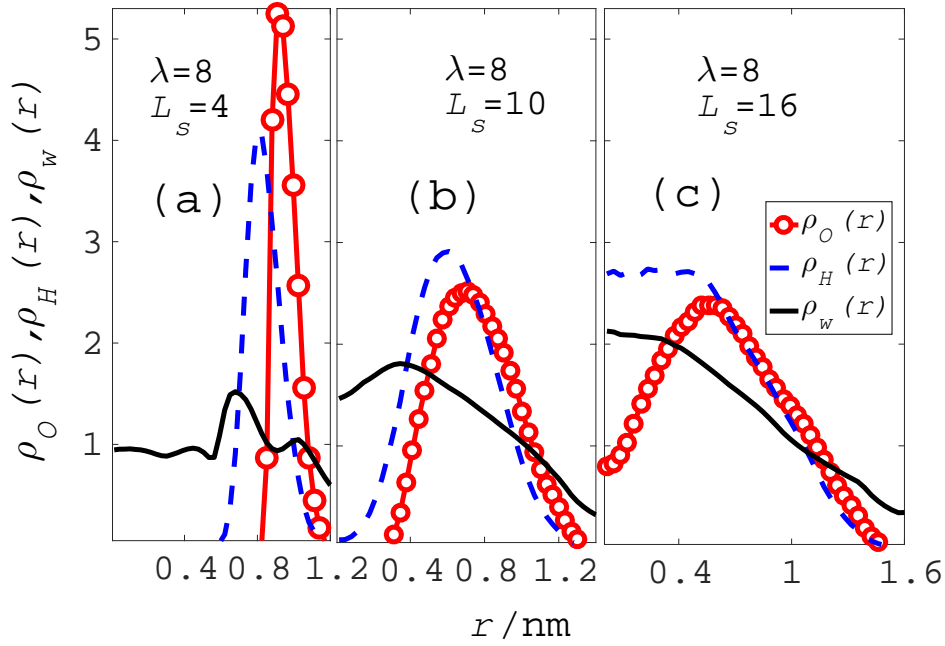}
\end{center}
\vspace{-0.7cm}
\caption{(Color online) 
Normalized radial distribution
  functions $\rho_j(r)/\rho_j^0$ for the 
{\need{ 
 hydronium 
}}
ions ($j$=$H$), water molecules ($j$=$w$) and 
  the sulfonate oxygens ($j$=$O$), 
as a function of the radial distance $r$ from  the pore center
 for the simulation runs C1 (a), C3 (b), and C6 (c) from
   Table~\ref{table-abc}.   The normalizing
  factor $\rho_j^0=N_j/V$, where $V=\pi d^2 L/4$ is the pore volume,
  and $N_j$ is the  number of particles of the sort $i$. 
Full red lines with circles- sulfonates, dashed blue lines- ions, 
solid black lines- water molecules.
 \label{fig-13-new}}
\end{figure}

In the {\bf case A} and 
{\bf B} simulations discussed in previous sections,  
the variation of $L_s$ and $d$ were accompanied with change 
in the water content $\lambda$. In the current section 
we  fix the water content to  $\lambda$=8, the average 
hydration number for Nafion-like ionomers, 
and focus on finding an enhanced  ion diffusion in the charged pores. 
We choose the run B2 as a reference run for the {\bf case C} simulations, 
and vary the sidechain protrusion length $L_s$ from 4 monomers to 16 monomers.
Simulation results for the radial density distribution 
$\rho_j(r)$ of the ions ($j$=H), 
             sulfonate oxygens ($j$=O), and the 
        water molecules ($j$=w),  
shown in Figure~\ref{fig-13-new}, resemble the radial 
distributions for the {\bf case B} runs in Figure~\ref{fig-10-new} \cite{explain}. 
Here, again, in the case of long sidechains with $L_s$=16 monomers, 
  the ions are electrostatically  delocalized in the central pore  
area and generate a radial charge separation in the pore.  
A remarkable fact is the flattening of the ion 
 and water profiles in the central pore area in  Figure~\ref{fig-13-new}c at 
$\lambda$=8, in contrast to the steep gradients  of the $\rho_H(r)$
and $\rho_w(r)$ in the same area as seen in  Figure~\ref{fig-10-new}c for $\lambda$=2.8. 
At the same time,  no such flattening  of the  
sulfonate profile in Figure~\ref{fig-13-new}c from the increased 
water content is visible. 
As a result of these  observations, we conclude that 
the radial charge separation in the pore decreases if 
more water is absorbed in the pore.

\begin{figure}[!ht]  \hspace{2cm}
{\Large \bf(a)}   \hspace{4.5cm} {\Large \bf (b)}   \hspace{5cm}  {\Large \bf (c)}   \\
 \hspace{1.0cm} $\lambda$=8, $L_s$=4  monomers  
 \hspace{2cm}   $\lambda$=8, $L_s$=10 monomers  
 \hspace{2cm}   $\lambda$=8, $L_s$=16 monomers 

\includegraphics*[width=0.28\textwidth,height=0.57\textwidth]{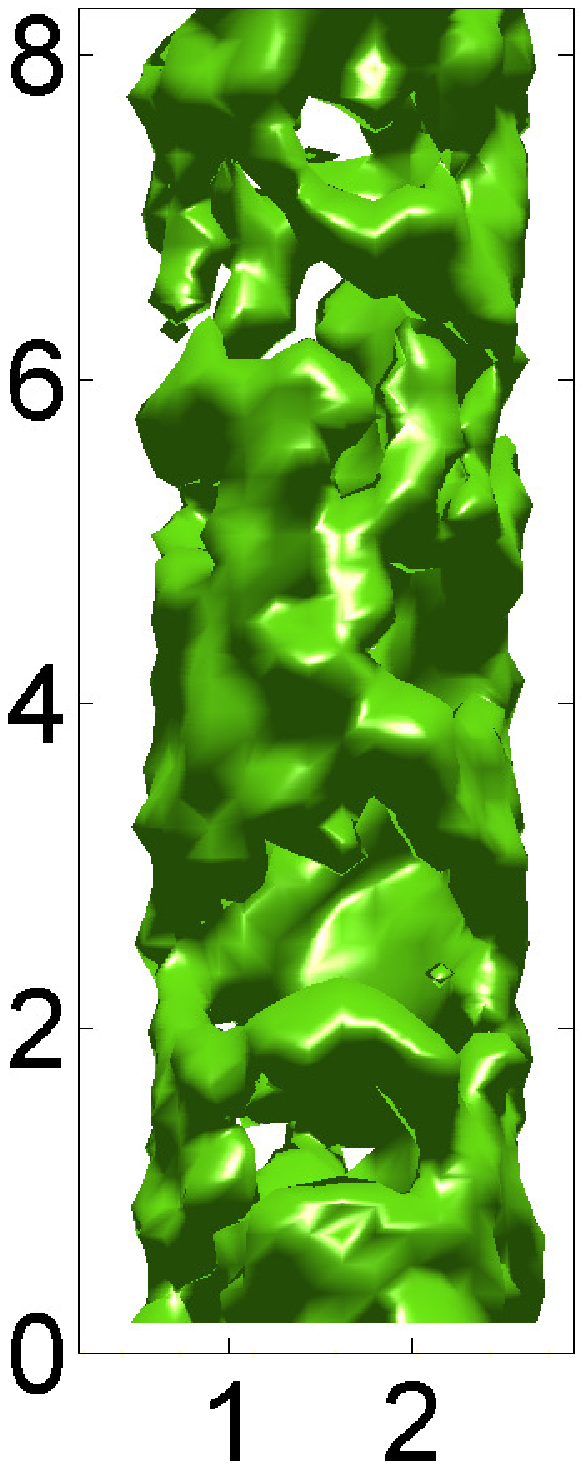} 
\includegraphics*[width=0.30\textwidth,height=0.57\textwidth]{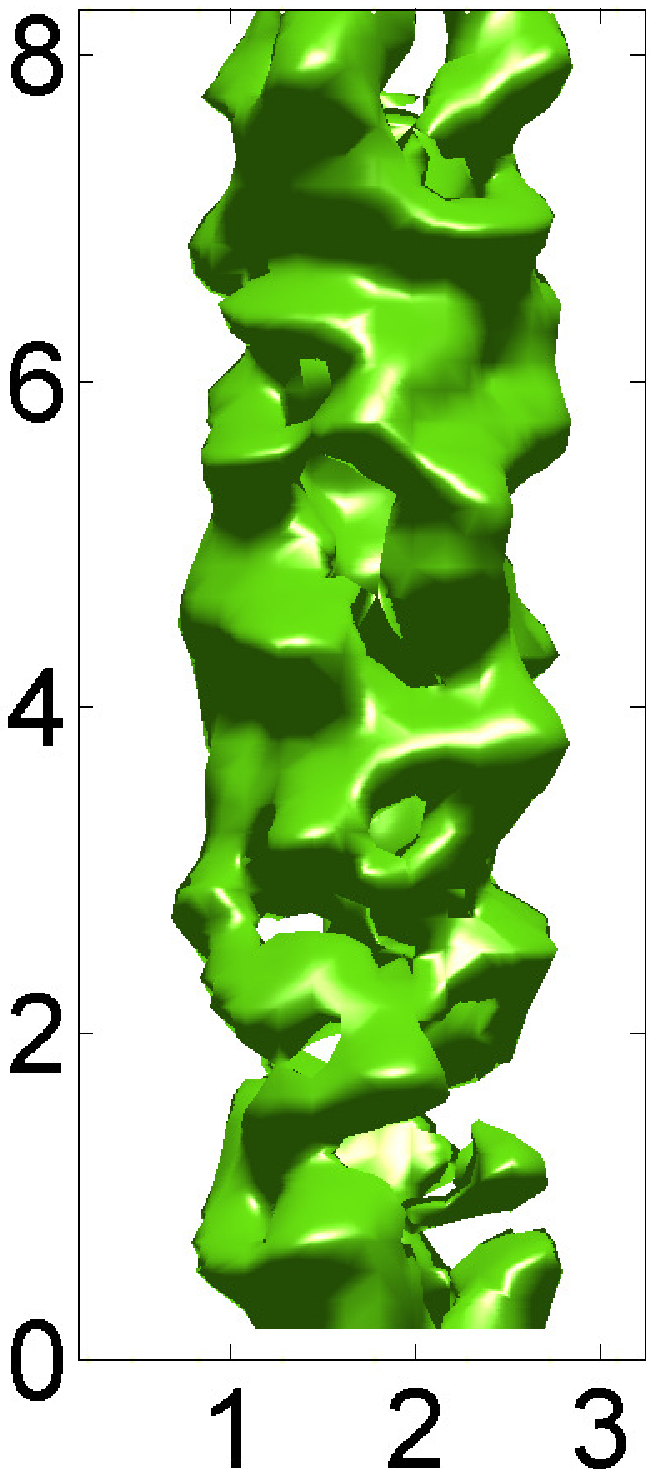}
\includegraphics*[width=0.33\textwidth,height=0.57\textwidth]{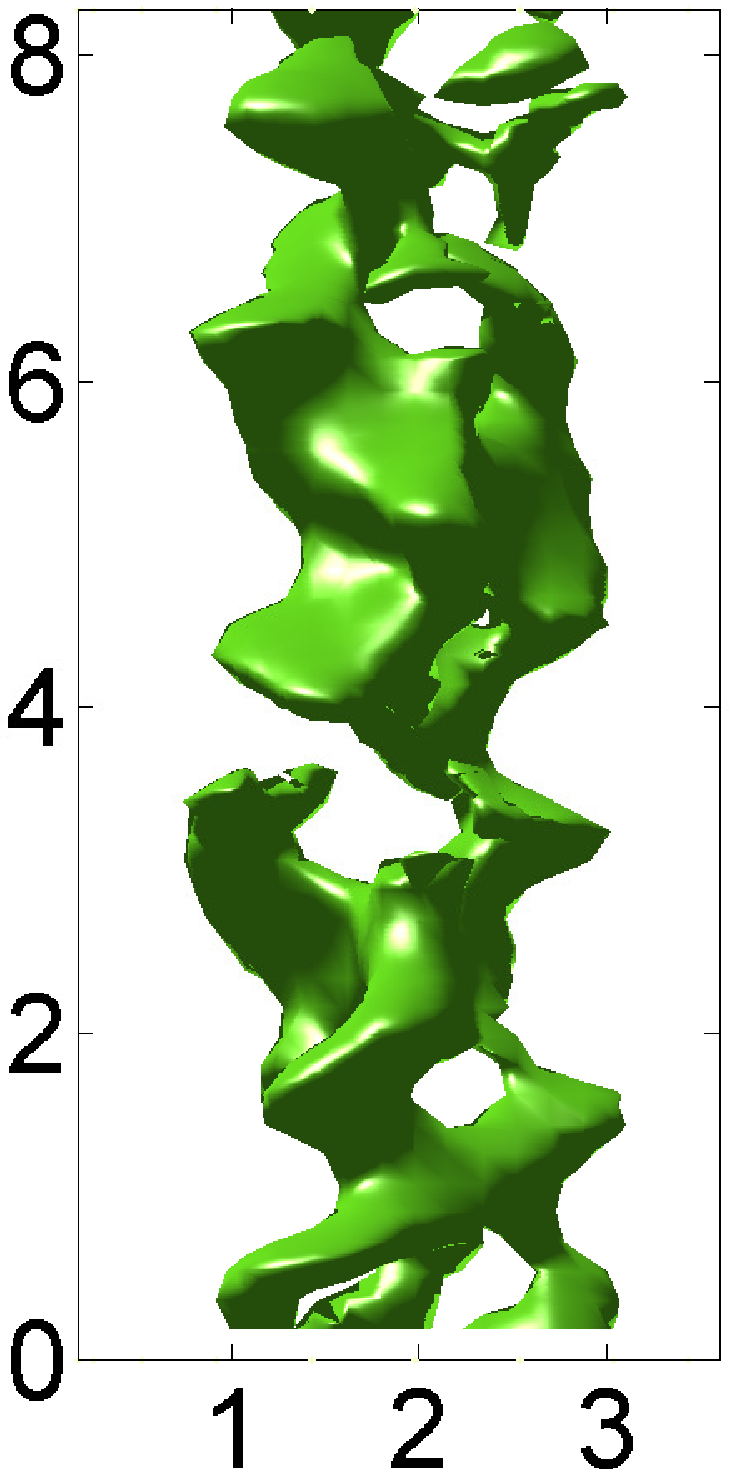}
\includegraphics*[width=0.28\textwidth]{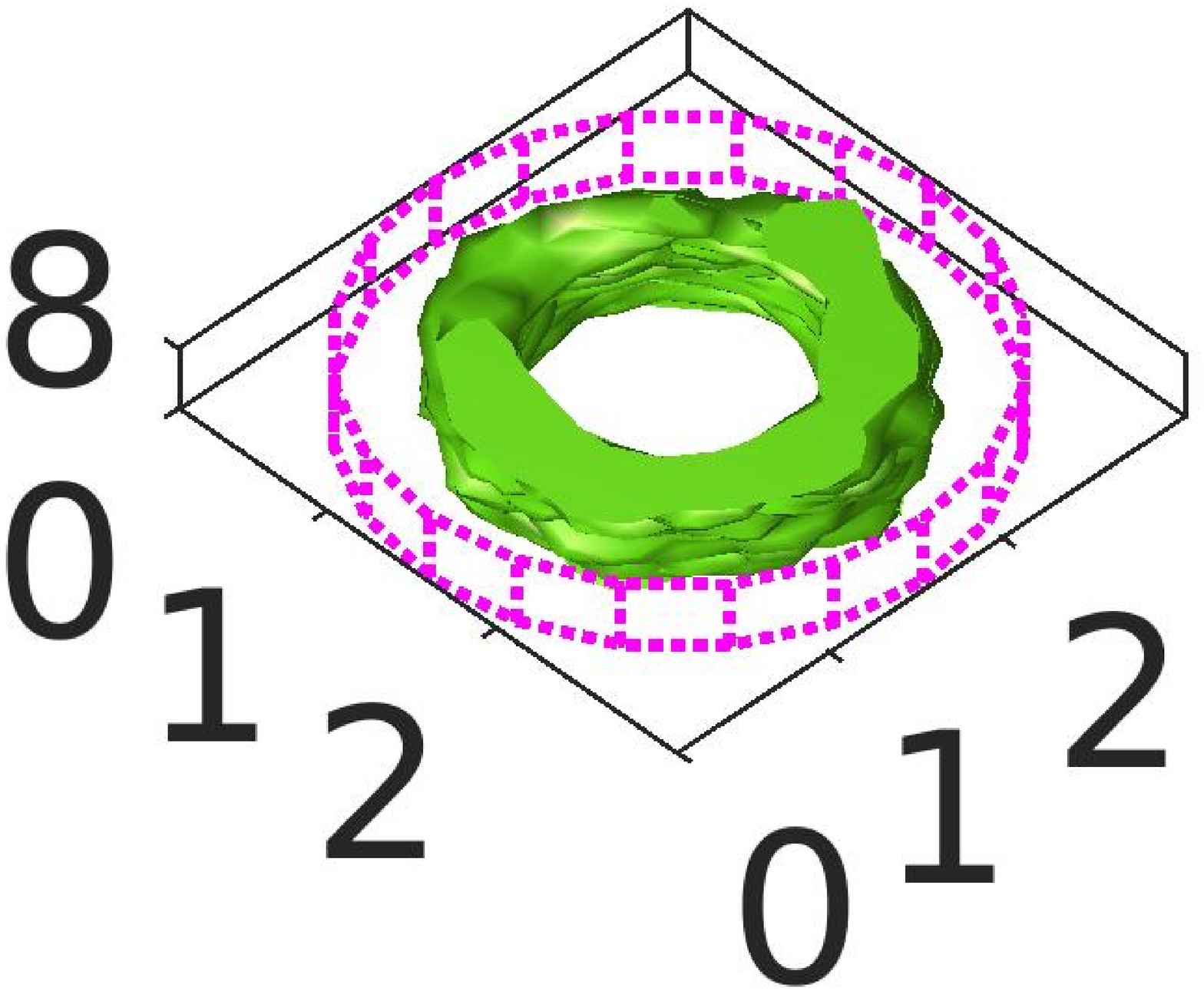}
\includegraphics*[width=0.30\textwidth]{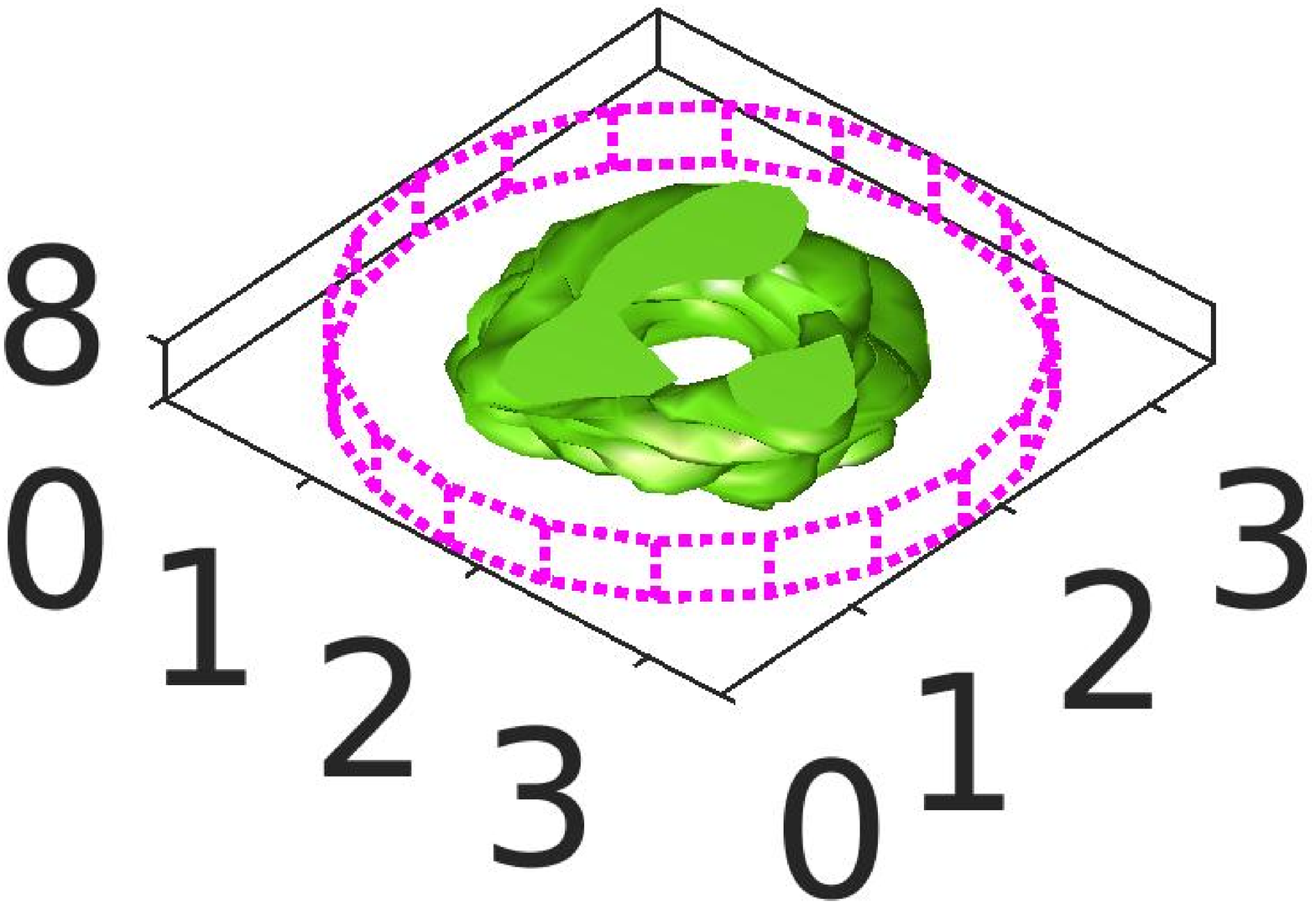}
\includegraphics*[width=0.33\textwidth]{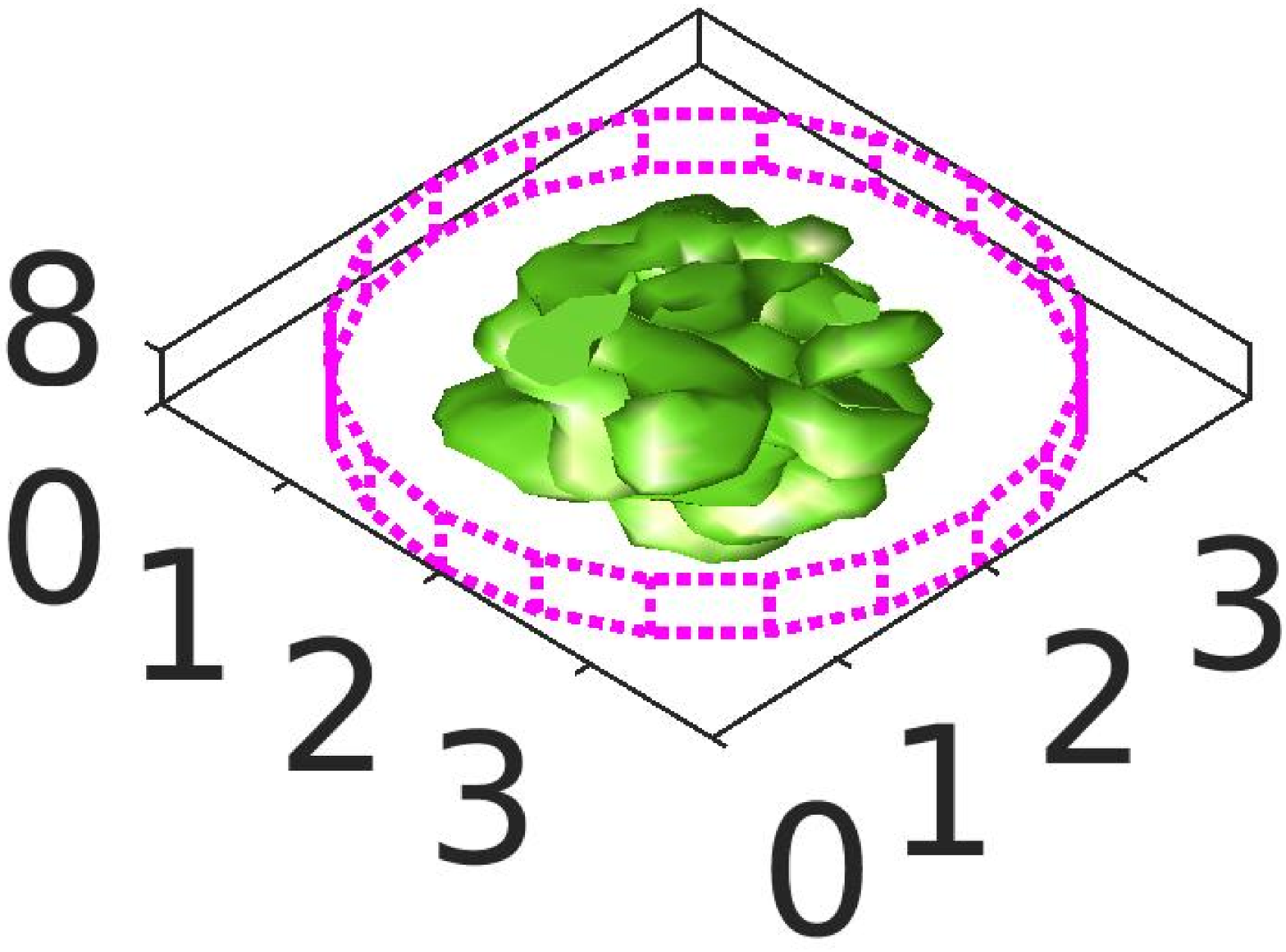}
\caption{(Color online)   Three dimensional ion  density distribution 
$\rho_H(\vec r)$ for the simulation runs C1 (a), C3 (b), and C6 (c) from
   Table~\ref{table-abc}. 
The figures on the upper row correspond to  
the projection of the density distribution 
on the $xz$-plane (only a section of the pore is shown).   
The figures on the bottom  row correspond to
the  projection of the density distribution 
on the $yz$-plane with a slight angular tilt for showing the internal
surface of the density distribution. The pink-colored circles and lines in the 
bottom row figures represent the position of the pore  surface. 
 \label{fig-14-new}}
\end{figure}

From the similarity of the ion and sulfonate radial 
distribution profiles in   
Figure~\ref{fig-13-new}b and 
Figure~\ref{fig-10-new}b  it is more likely to expect that  
the run C3 with $L_s$=10 monomers and $d$=3.3 nm would  provide maximal ion diffusion 
in the pore. The 3D distribution of ions for some of the {\bf case C} runs
are plotted in  Figure~\ref{fig-14-new}.
{\need{
 The averaging of the ion positions in these plots was done
during the 50 fs long simulations in fully equilibrated system to get
 instantaneous   ion distribution profiles in the pore.
}}
It is seen that for a shorter
sidechains with $L_s$$\leq$10 monomers, 
the ions mostly accumulate near the pore walls, where  
they form  continuous hollow-cylinder-like ionic pathways. The hollow 
area of this structure is filled with water molecules. 
For the sidechains with bigger monomer numbers $L_s$,  the ions accumulate at the pore center by forming  
larger clusters with disrupted connections along the pore axis. 
These disrupted and  hollow areas  are filled either 
with water molecules or with  sulfonates, with the  latter hindering
 smooth ion passage through the pore.

\begin{figure}[!ht]
\begin{center}
\includegraphics*[width=0.63\textwidth,height=0.59\textwidth]{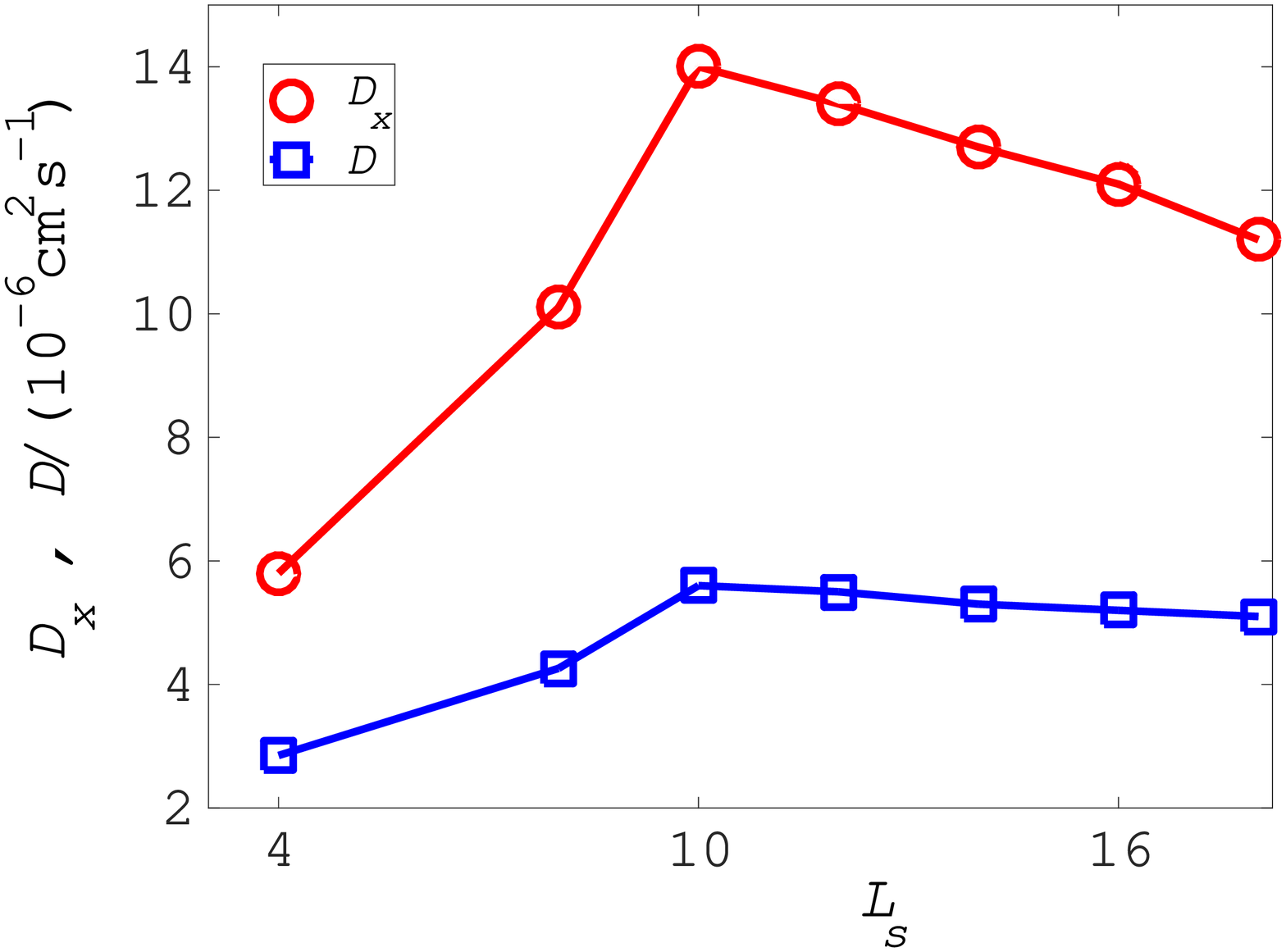}
\end{center}
\vspace{-0.75cm}
\caption{(Color online) 
The total ion  diffusion coefficient $D$ 
(blue line with squares), and the 1D ion diffusion 
coefficient $D_x$ along the pore longitudinal axis $x$ 
(red lines with circles)
for  the runs C1-C7 from  Table~\ref{table-abc}. 
The number of grafted sidechains is  $N_s$=200. 
 \label{fig-15-new}}
\end{figure}

Calculated diffusion coefficients for the ions as a 
function of the sidechain  protrusion length 
$L_s$ for the runs C1-C7 are plotted in Figure~\ref{fig-15-new}. 
The axial diffusion coefficient $D_x$ for the entire range 
of the sidechain  protrusion  length $L_s$ variation is more than twice 
as large as the full 
diffusion coefficient $D$ of the ions, a result of the strongly suppressed ion diffusion
in the radial direction, perpendicular to the pore walls. A maximum in the 
$D_x$ is seen for  the sidechain  protrusion  
length $L_s$=10 monomers (run C3), as is expected from 
the radial distributions of ions and sulfonates in  Figure~\ref{fig-14-new}.
As will be  discussed in the next section, it is indeed possible 
the predict the pore parameters at which  maximal ion diffusion 
along the pore axis is expected.

\begin{figure}[!ht]
\begin{center}
\includegraphics*[width=0.7\textwidth]{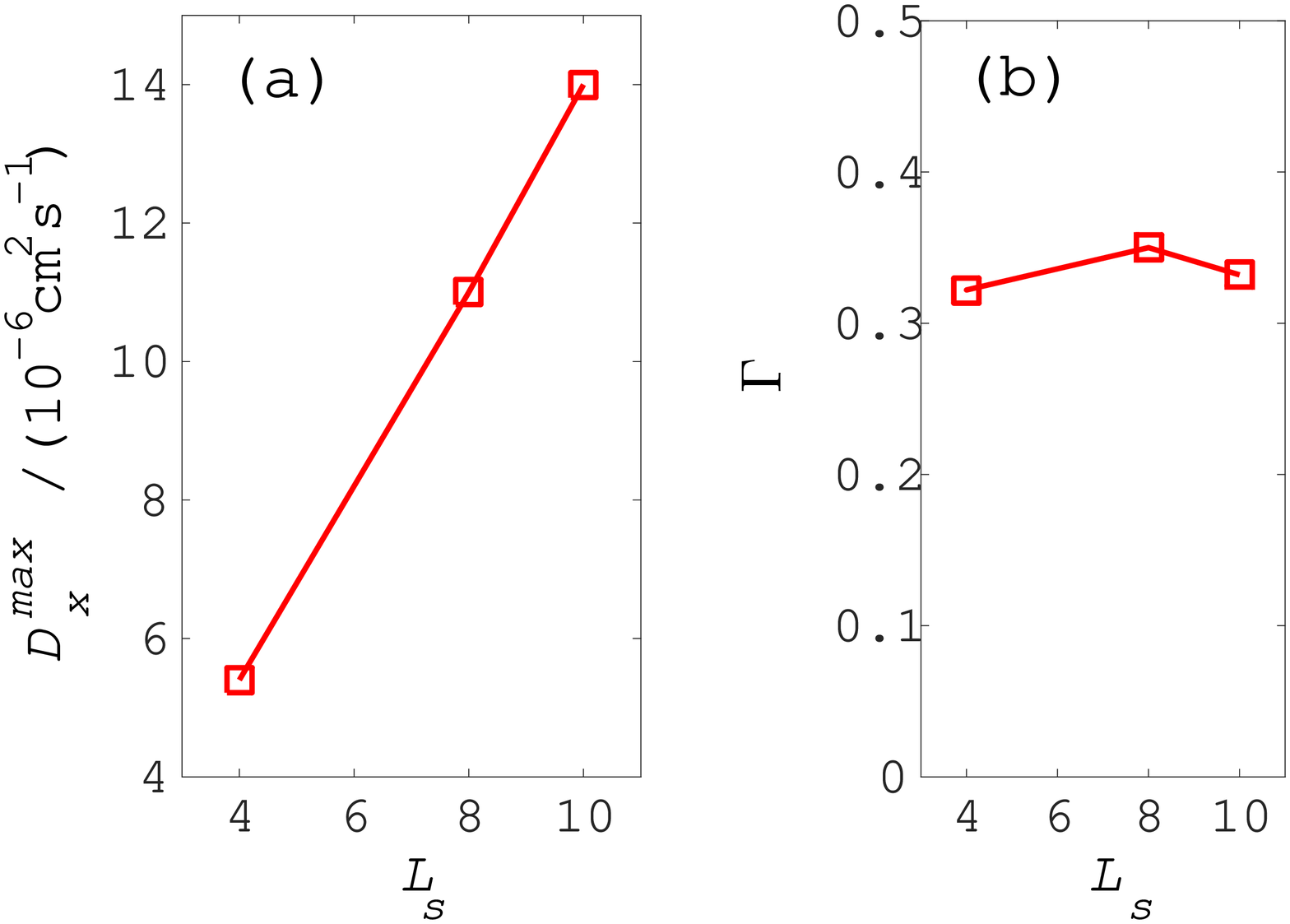}
\end{center}
\vspace{-0.7cm}
\caption{(Color online) 
(a) The dependence of the maximal ion diffusion 
$D_x^{max}$ from the runs A3, B4, and C3, on the sidechain protrusion 
length $L_s$. (b) Scaling parameter $\Gamma$  for the pore parameters $d$ and $L_s$
from Eq.(\ref{gamma}) with the Flory exponent 1/4. 
 \label{fig-16-new}}
\end{figure}

{\mod{ 
\subsection{Maximal ion diffusion in charged pores}
}}
\label{maximal-diffusion}
In this section we collect and discuss the maximal ion diffusion coefficients $D_x^{max}$ 
along the pore axis, detected for the runs A3, B4, and C3.  The dependence of the $D_x^{max}$  on the sidechain
protrusion length $L_s$ is plotted in Figure~\ref{fig-16-new}a.
It appears that  $D_x^{max}$ is a linear function of the sidechain protrusion 
length $L_s$.
Taking into account the fact that long sidechains in larger pores behave like 
short sidechains in narrow pores, it seems plausible 
to find a scaling rule for the pore parameters $d$ and $L_s$ at which the ion diffusion is maximal along
the pore axis. For the average sidechain extension into the pore center 
it is reasonable to introduce the  gyration radius of the sidechain 
\begin{equation}
R_g=b L_s^{\nu}
\end{equation}  
with a phenomenological Flory-like exponent $\nu$.
Here $b$ is the segment length of the sidechain, $b$=0.154 nm for 
Nafion-like ionomers,
For the polymer chains  with excluded-volume interactions the exponent varies 
between 1/2 and 3/5   \cite{flory-exponent}, whereas for attractive coil-globule collapsing
polymers the Flory exponent is 1/3. A lower value  $\nu$=1/4    
 has been reported for dilute branched polymers 
\cite{lubensky-1980-flory-exponent,joanny-1980-flory-exponent}. 
Twice the  gyration radius can be considered as the protrusion length of the 
sidechain into the pore center, $l_s$=$2 R_g$.

{\need{
 In Figure~\ref{fig-16-new}b we test a phenomenological fit 
\begin{equation}
\Gamma = \frac{2 l_s}{d} =  \frac{4 R_g}{d}  
       = \frac{4}{d} \, {b  L_s^{\nu}}
\label{gamma}
\end{equation} 
for the pore parameters $d$ and  $L_s$  at which the ion diffusion is maximal.
The scaling parameter $\Gamma$ measures the ratio of the sidechain protrusion length $l_s$  
to the pore radius $d/2$. We found that
for the fitting exponent $\nu$=1/4 the pore scaling parameter is almost 
constant at $\Gamma$$\approx$0.33 for $L_s$ changed from 4 monomers  to 10 monomers, 
and $d$ changed from 2.7 nm to 3.3 nm. 
The fitting value $\nu$=1/4 is the same as the Flory exponent for branched polymers. This 
is a reasonable conclusion, because the sidechains in the pore 
create clusters which indeed can be viewed as 
a branched polymer. 
The result $\Gamma$$\approx$0.33 implies that the sidechain extension
$l_s$ into the pore center should be about 1/3 of the pore radius for 
the pore to provide  maximal ion diffusion along the pore axis. 
}}

{\need{ 
An interesting question is how the optimal protrusion length $l_s$ depends on the 
diffusing ion size $\sigma_i$. Apparently there is no direct link between
these two quantities. Under the condition that the free volume for the ions in the pore is kept constant, 
the change in $\sigma_i$ will affect the ion diffusion rates in 
all simulated cases A, B, and C in the same manner. Smaller ions will be more attracted to the 
oppositely charged sulfonates which will slow down their diffusion, whereas larger ions will be more 
readily released from the sulfonates and thus have higher diffusion rates. Therefore, we think that 
the diffusion lines  shown in Figures 8, 12, 15, and 
16a will retain their shape but shift either up or down depending on 
the decrease or increase in  $\sigma_i$ respectively.    
}}

{\need{ 
The maximal diffusion coefficients $D_x^{max}$ shown in Figure~16 are in the range 
$D_x^{max}$$\approx$(10--14)$\times$10$^{-6}$cm$^2$/sec for the   
sidechain lengths $L_s$$\approx$8--10 monomers. These values 
 are about half the size of 
 the self-diffusion of water molecules in bulk water
 $D_{H_2O}$$\approx$23$\times$10$^{-6}$cm$^2$/sec   
\cite{perrin-2007,choi2005,kerisit-2009,kreuer-2004-review,krynicki-1978-water-self-diffusion,agmon-1995-water-grotthuss}, 
and about 30--40\% smaller than the simulated vehicular (en-masse) diffusion values for hydronium ion in bulk water
    $D_{H_3O}$$\approx$(17--20)$\times$10$^{-6}$cm$^2$/sec \cite{li-2001-partial-charges,choi2005}.  
Obviously, it is  more relevant  to compare our data for $D_x^{max}$ with the water and  hydronium diffusion 
coefficients in ordinary ionomers with $L_s$$\approx$8 monomers
 like the Nafion membrane. In this case, however,  one has to distinguish 
between the local $D^l$ and long-range $D^{lr}$ diffusion coefficients, with the $D^l$ corresponding to the particle 
diffusion in a single cluster at shorter time scales, and 
$D^{lr}$ corresponding to the particle diffusion at longer time scales when the particle has enough time to
pass through several hydrophilic clusters in the ionomer. Generally, 
$D^{l} \gg D^{lr}$ because of the tortuousity and narrow bridges between neighboring 
hydrophilic clusters in the ionomer \cite{zawod-1995}. 
For example,  in a  fully hydrated ionomer with $\lambda$$\approx$15--17 the water molecule has a local self-diffusion coefficient  
  $D_{H_2O}^l$$\approx$13$\times$10$^{-6}$cm$^2$/sec  which is about twice 
as large as  its  long range self-diffusion coefficient $D_{H_2O}^{lr}$$\approx$6$\times$10$^{-6}$cm$^2$/sec  
 \cite{perrin-2006},  and much higher than its  long range self-diffusion coefficient
  $D_{H_2O}^{lr}$($\lambda$$\approx$5--10)$\approx$(1--5)$\times$10$^{-6}$cm$^2$/sec 
in partly  hydrated Nafion \cite{eikerling-2008,cui-2008}.
Additionally, one has to bear in mind that 
experimentally determined diffusion coefficients for hydronium ions  
include the Grotthuss proton hopping mechanism,   which is not accounted for in our simulations. 
 For example, $D_{H_3O}^{lr}$ in  fully hydrated Nafion with the Grotthuss hopping accounted for reaches  
 $D_{H_3O}^{lr}$(Grotthuss + en-masse)$\approx$(12--14)$\times$10$^{-6}$cm$^2$/sec \cite{li-2001-partial-charges}
which is more than twice as large as  the en-masse diffusion  
 $D_{H_3O}^{lr}$(en-masse)$\approx$(4--5)$\times$10$^{-6}$cm$^2$/sec \cite{li-2001-partial-charges,sun-2015-md-simulation-diffusion}. 
The same is valid in partly hydrated Nafion with $\lambda$$\approx$6--10, where 
 $D_{H_3O}^{lr}$(Grotthuss + en-masse)$\approx$(2--5)$\times$10$^{-6}$cm$^2$/sec 
is larger than 
 $D_{H_3O}^{lr}$(en-masse)$\approx$(1--2)$\times$10$^{-6}$cm$^2$/sec \cite{perrin-2007,cui-2008}.
}}

{\need{
Our data for $D_x^{max}$ corresponds to the long-range diffusion $D^{lr}$ of hydronium ions without the 
Grotthuss contribution. Therefore,  we get 
a diffusion rate 2--3 times larger  than  $D_{H_3O}^{lr}$(en-masse) 
in fully hydrated 
Nafion \cite{li-2001-partial-charges,sun-2015-md-simulation-diffusion}, 
and about 5--7 times larger  in partly hydrated Nafion \cite{perrin-2007,cui-2008}. 
}}

{\mod{ 
\section{Conclusions}
}}
\label{conclusons}

In this work we analyzed  ion diffusion phenomena in charged nanopores grafted 
with ionomer sidechains. Our aim was to  determine the  optimal pore parameters
 for obtaining   enhanced ion diffusion.  
{\need{ 
We found that in the case of short sidechains, the hydronium ions mostly occupy the pore wall area 
and  their distribution is strongly disturbed by the electric field 
of sulfonates in swollen pores. Additionally, the hydronium ions delocalize from their host sulfonate 
groups in wide pores. However, an increase in the anchoring distance $r_{ss}$ and associated with it  
a stretching of sidechains  restricts the  ion diffusion in swollen pores. 
}}
In narrow pores, on the other hand,  the structuring and polarization effects  of the  water molecules
 hinder  the free movement of the ions in the pore. 
Consequently, the ion diffusion in the charged pores grafted with short sidechains
  becomes  a nonlinear function of the pore diameter. 
It attains a  maximum in the pores with less structured water,  
 high surface charge density, and  moderately delocalized ions occupying only the 
pore wall area. 

In the charged pores with  a fixed pore size $d$, according to our simulations, 
the clustering of the sulfonates is strongly regulated
 by their protrusion length $L_s$ rather than by
the  water content $\lambda$.  
For the longer $L_s$, when the sidechain tips reach the pore center,   
a radial charge separation occurs in the pore. The pore center 
with excess ions is  charged positively, while the pore  wall area 
with excess sulfonates is charged negatively. Such  
 charge separation, which is associated with 
the sulfonate clustering in the pore central area,
 suppresses the ion diffusion along the pore axis. 
 An enhanced ion diffusion was found in  
the pores grafted with  medium-size sidechains provided that 
the  ions do not enter the central pore area, and the water is less structured 
around the ions. 
A similar conclusion is also made for the pores with a fixed water content. 
The medium-sized sidechain with tips not entering  the pore central area 
  allows the ions to form  hollow-cylinder type hydrophilic pathways in the 
channel. The existence of such cylindrical shells with smooth 
and uninterrupted 
ion pathways is the necessary condition for getting  
an easy  ion passage along the pore axis.

By collecting and analyzing the pore parameters for which our simulations have detected 
 maximal ion diffusion along the pore axis, we found that the
 ion diffusion $D_x^{max}$ has a linear dependence on the sidechain 
protrusion length $L_s$. The long chains, provided that they do not 
reach the pore center, have very flexible tips which assist the 
ion diffusion along the pore axis.  We also proposed a simple scaling rule for the pore 
parameters $d$ and $L_s$ with a Flory-like exponent 1/4. It appears that 
a maximal ion diffusion along the pore axis is possible if the 
effective length of the sidechain extension into the pore center,
measured as twice the gyration radius of the sidechain with 
the Flory-like exponent 1/4, is about 1/3 of the pore radius $d/2$.

{\need{ 
The simulated  axial diffusion coefficients have maximal values 
in the range $D_x^{max}$$\approx$(10--14)$\times$10$^{-6}$cm$^2$/sec for  
 water content $\lambda$$\approx$6--8, and 
sidechain length $L_s$$\approx$8--10. These diffusion rates  are 
about  2--3 times larger  than the diffusion rate  $D_{H_3O}^{lr}$(en-masse) in fully hydrated 
Nafion \cite{li-2001-partial-charges,sun-2015-md-simulation-diffusion}, 
and about 5--7 times larger than the diffusion rate  $D_{H_3O}^{lr}$(en-masse) in partly hydrated Nafion 
\cite{perrin-2007,allahyarov-2011-diff-archit,cui-2008}. 
We believe that experimental realizations of our grafted pore set-up will produce even higher ion diffusion rates because of the 
 Grotthuss structural diffusion contribution to $D_x^{max}$.
For example, the Grotthuss proton diffusion rate in bulk water is 
$D_p$(Grotthuss)=70$\times$10$^{-6}$cm$^2$/sec, and with the water 
self diffusion $D_p$(en-masse)=23$\times$10$^{-6}$cm$^2$/sec added to it,  
the total proton diffusion rate roughly increases up to   
$D_p$(Grotthuss + en-masse)=93$\times$10$^{-6}$cm$^2$/sec  \cite{choi2005,agmon-1995-water-grotthuss}. 
In Molecular Dynamics simulations the Grotthuss hopping can be accounted for by implementing  the  empirical valence bond method
\cite{eikerling-kornyshev-2001,paul-paddison-2005-pore,walbran-2001}.
}}

These results obtained for ion diffusion in charged pores 
may be helpful in the  further 
development of  new emerging technologies, such as  energy storage in 
nanoporous metals \cite{zhang-2012-nanoporous-metals}, or  
a new class of soft fluid actuators \cite{sritharan-2016-liquid-actuators}. 
Our results also might contribute to the development of a new class of elastomer 
actuators called {\it metallic muscles}  \cite{detsi-2013-charged-pore-actuator}.
In these applications the pores of the metal inclusions  are polymer coated
 and charged by doping with sulfuric acid. Nanocomposite elastomers with these nanoporous metal 
inclusions,   besides generating electrostrictive strains 
\cite{allahyarov-actuator-2015-multilayer,allahyarov-electroactuator-2015-inclusions,allahyarov-actuation-2016}, will also generate additional strain from the 
interaction of the electrostatic double layers between these highly charged objects. 
Then the total actuation response of the nanocomposite to the applied field will be 
greatly enhanced.

We also propose  that the  concept of the  self-assembling
of  amphiphilic diblock copolymers 
in cylindrical confinements \cite{xiang-2004,srinivas-2004}
can be used to obtain grafted pores. In this case the free hydrophilic ends of the polymer
will form brush conformations \cite{dimitrov-2006,kouts-2009}. 
The functionalization and hydration of these brushes will turn the pore 
into an ion-conducting channel.

\vspace{0.5cm}
\section*{Acknowledgments}
E. Allahyarov is thankful for support of this work by 
the Deutsche Forschungsgemeinschaft (DFG)  through the grant 
AL 2058/1-1 on nanocomposite elastomer actuators. 
H. L\"owen thanks the Deutsche Forschungsgemeinschaft
 for support of the work through the SPP 1681 on magnetic hybrid materials.
P. Taylor acknowledges support by the 
American Chemical Society Petroleum research Fund under Grant PRF\#51995-ND7.


\begin{thebibliography}{00}
\bibitem{explain}
The  Figure~\ref{fig-13-new}a for the run C1 and 
the  Figure~\ref{fig-10-new}a for the run B2 
are the same.
\end{thebibliography}
\end{document}